\documentclass[twocolumn,useAMS,usenatbib]{mn2e}
\usepackage{graphicx,amssymb,amsmath}
\usepackage{bm,txfonts}
\usepackage{color}

\newcommand{\rmd}{{\rm d}}
\newcommand{\rme}{{\rm e}}
\newcommand{\rmi}{{\rm i}}

\newcommand{\beq}{\begin{equation}}
\newcommand{\eeq}{\end{equation}}
\newcommand{\beqa}{\begin{eqnarray}}
\newcommand{\eeqa}{\end{eqnarray}}

\newcommand{\cmnt}[1]{}

\title[Dispersion of incoherent sources]{Inference of dispersion measure from incoherent time-steady sources}

\author[Hirata \& McQuinn]
 {Christopher M. Hirata$^{1}$ \& Matthew McQuinn$^{2}$
\\$^1$Department of Physics and Department of Astronomy, The Ohio State University, 191 West Woodruff Avenue, Columbus, Ohio 43210, USA
\\$^2$Department of Astronomy, University of California, Berkeley, California 94720, USA; Hubble fellow}

\date{11 December 2013}

\begin{document}
\maketitle

\begin{abstract}
Several recent papers have proposed schemes by which a dispersion measure, and hence electron column, could be obtained from a time-steady, incoherent radio source at a cosmological distance (such as an active galactic nucleus).  If correct, this would open a new window on the distribution of intergalactic baryons. These schemes are based on the statistical properties of the received radiation, such as the 2- or 4-point correlation function of the received electric field, and in one case on the quantum nature of the electromagnetic field. We show, on the basis of general principles, that these schemes are not sensitive to dispersion measure (or have an extremely small signal-to-noise ratio), because (i) the classical 2-point correlation function is unaffected by dispersion; (ii) for a source with a large number of incoherently emitting electrons, the central limit theorem obliterates additional information in higher-order functions; and (iii) such an emitter produces a radiation density matrix that is equivalent to a statistical distribution of coherent states, which contains no information that is not already in the statistics of the classical waveforms.  Why the proposed observables do not depend on dispersion measure (or have extremely tiny dependences) is discussed in detail.
\end{abstract}


\section{Introduction}
\label{sec:intro}

Estimates of dispersion measure (DM) using the frequency-dependent arrival time of pulses from pulsars have provided some of the most useful constraints on ionized gas in the interstellar medium of the Milky Way Galaxy \citep{1993ApJ...411..674T, 2002astro.ph..7156C} and in the Galactic halo \citep[e.g.][]{2013ApJ...762...20F}.  These estimates provide a direct measurement of the electron column density to each source, $\int n_e\,\rmd\ell$.  They are independent of clumping factors, gas temperatures, and (in some wavebands) extinction, in contrast to emission-based probes of ionized gas.  They are also independent of the metallicity and ionization state, unlike absorption line tracers.  It would be extremely useful to have a similar electron column density probe for the intergalactic medium \citep[e.g.][]{2013arXiv1309.4451M}.

Unfortunately, pulsars are intrinsically faint and the most distant ones observed thus far are in the Magellanic Clouds; for the foreseeable future, pulsar-based DMs are unmeasureable at cosmological distances. Two recent developments have renewed interest in dispersion measure as a cosmological probe. One has been the discovery of radio transients with durations of order milliseconds or less with a frequency-dependent arrival time consistent with plasma dispersion (\citealt{2013Sci...341...53T}; see also \citealt{2007Sci...318..777L}). These transients' implied DM is far greater than that expected from the Milky Way and is consistent with an origin at cosmological distances ($z$ of order unity).  However, a terrestrial origin has not been excluded \citep[see e.g.][]{2011ApJ...727...18B, 2013arXiv1310.2419L}. The other major development -- and the subject of this paper -- is the suggestion that DM could be measured in a {\em continuous} rather than pulsed source \citep{2013MNRAS.433.2275L, 2013ApJ...763L..44L}. If correct, then dispersion measures could be obtained using radio-loud active galactic nuclei (AGNs) as backlights, opening a completely new window on the study of intergalactic gas.

There are two recently proposed schemes for measuring the DM to a continuous synchrotron source. One uses the 2-point correlation function of the electric field between two neighbouring frequency channels. The idea is that in the presence of dispersion, these channels' cross-correlation should peak at a nonzero lag, with the signal arriving first in the higher-frequency channel \citep{2013MNRAS.433.2275L}. The other scheme is to use the 2-point correlation function of intensity fluctuations (which is a 4-point correlation function of the received electric field).  It relies on the intuition that since dispersion within a frequency channel smears out the received synchrotron pulse from each electron, the timescale at which the intensity fluctuates should be increased relative to the DM$=0$ case, where the correlation time is roughly the inverse of the bandwidth, $\sim\Delta f^{-1}$ \citep{2013ApJ...763L..44L}.

This paper shows that no scheme for measuring the DM to a time-steady incoherent radio source can work. The conceptual basis for this result is that the 2-point correlation function of the electric field contains no phase information and, hence, is insensitive to the phase shift that is produced by dispersion.  While the non-Gaussian (or connected) part of higher-order statistics, such as the 3- and 4-point correlation functions, are indeed sensitive to dispersion measure, we show that the non-Gaussian part is undetectable for any potential incoherent source.  This is a consequence of the the central limit theorem: the number of electrons contributing to a source's flux at any time is large and, hence, the signal is very Gaussian.  There are exceptions to this argument:  (1) the relative phase of both polarizations is observable in the 2-point correlation function (which is why rotation measure is measurable), and (2) the relative phase of the received wavefront is altered both spatially or temporally by density gradients along the sightline (yielding scintillations). However, neither of these loopholes is relevant to the determination of the dispersion measure, which is polarization-independent, does not vary with observer position over accessible baselines, and does not produce amplitude fluctuations.

%This paper will present the mathematical demonstration that DM cannot be measured from time-steady incoherent radio sources. 
This paper will also explain why the proposed observables in \citet{2013MNRAS.433.2275L} and \citet{2013ApJ...763L..44L} do not depend on DM, despite it seeming plausible that they would. In the case of the former observable (the 2-point correlation function between adjacent bandpasses), in addition to the effect discussed in \citet{2013MNRAS.433.2275L} where an electronic pulse arrives earlier in the higher band, dispersion also distorts the pulse shape.  We show that this distortion correlates the beginning of the lower-frequency signal with the end of the higher-frequency signal, exactly canceling the impact of the delay on observed correlations. In the case of the latter observable (the temporal correlation of intensity fluctuations, or the 4-point correlation function of the electric field), for dispersed electron pulses, some electrons contribute correlated fluctuations in the intensity at unequal times -- i.e. they contribute positively to $\langle \delta I(t_1) \delta I(t_2) \rangle$ -- but some electrons contribute anti-correlated fluctuations. The net result is that the intensity fluctuations decorrelate on timescales longer than $\sim\Delta f^{-1}$ even if the pulse from each electron is dispersed into a train whose temporal length is many times longer.

The only part of the intensity correlation function that does not decorrelate on timescales longer than $\Delta f^{-1}$ is the connected part, where all 4 electric fields are contributed by the same electron; we show that this connected part is suppressed by a factor of $1/N_e$, where $N_e$ is the number of electrons that contribute to the observed electric field at any given time. For any realistic AGN this is negligible. Recently, \citet{2013arXiv1311.1806L} claimed that the suppression is only by a factor of $1/\bar n_\gamma$, where $\bar n_\gamma$ is the photon occupation number seen by the observer. This would correspond to a quantum correction to the correlation function since it is proportional to Planck's constant: $1/\bar n_\gamma \sim hf/kT_{\rm sys}$. We show that quantum corrections do not enable one to measure the DM; the basic reason is that for near-classical sources, quantum intensity fluctuations (photon Poisson noise) are a feature of the uncertainty principle and its impact on the measurement process, and dispersion acts on the underlying wave function before the measurement is made.

The outline of this paper is as follows. In \S\ref{sec:stat-e}, we develop the formalism for correlation functions of the electric field received at a detector after processing through an arbitrary linear filter. We use this to prove in \S\ref{sec:stat-c} that the 2-point correlation function of the received electric field contains no information about the dispersion measure, regardless of the filters applied and which signal channels are correlated. In \S\ref{sec:ex}, the procedure suggested by \citet{2013MNRAS.433.2275L} is studied in detail as a special case.  % it is found that while the arrival time of the pulse from each electron is delayed in a lower-frequency channel relative to a higher-frequency channel, the modification of pulse shapes by dispersion leads to a peak in correlation function at zero lag regardless of DM. 
Intensity fluctuations are considered in \S\ref{sec:int-fluct}, where we show that the component that depends on dispersion is completely undetectable.  We further discuss the calculation in \citet{2013ApJ...763L..44L}. 
The possibility of masers as a source for DM measurements is briefly considered in \S\ref{sec:maser}.
We conclude in \S\ref{sec:disc}.
Appendix~\ref{app:QM} discusses the role of quantum mechanics in the emission process.

% Finally, \S\ref{sec:bispectrum} we estimate the maximum (but completely negligible) signal-to-noise that dispersion can be detected in a continuous, incoherent astrophysical source. -- REMOVED C.H.

\section{Formalism for electric field correlations}
\label{sec:stat-e}

Consider an optically thin incoherent synchrotron source. The notation here parallels that of \citet{2013MNRAS.433.2275L}. Index notation will be used as follows: capital Roman alphabet $ABC...$ denote signal channels (i.e. filtered electric fields in the chosen polarization state and filtered to the frequency band); lower case Roman indices $ijk...$ denote pulses from relativistic electrons in the source (the observed signal is the sum of the electric field from each such pulse); and Greek indices $\alpha\beta\gamma...$ denote polarization states. All summations are kept explicit. Fourier transforms here will use the normalization convention
\begin{equation}
E(t) = \int_{-\infty}^\infty \tilde E(f)\,\rme^{-2\pi\rmi ft}\,\rmd f
~~\leftrightarrow~~
\tilde E(f) = \int_{-\infty}^\infty E(t)\,\rme^{2\pi\rmi ft}\,\rmd t,
\end{equation}
and convolutions are denoted by $\ast$, with
\begin{equation}
[E \ast G](t) \equiv \int_{-\infty}^\infty E(t') G(t-t')\,\rmd t'
~~\leftrightarrow~~
[\widetilde{E \ast G}](f) = \tilde E(f) \tilde G(f),
\end{equation}
where tildes indicate the Fourier transform.

The electric field from the source as emitted is given by
\begin{equation}
E^{\rm em}_\alpha(t) = \sum_j {\cal E}_{j} \psi_{j\alpha}(t-t_j),
\label{eq:E}
\end{equation}
where the summation is over the pulse from each relativistic electron that contributes to the emission, ${\cal E}_{j}$ is a normalization of the source strength, $\psi_{j\alpha}(t)$ is the pulse profile, and $t_j$ is the time of the $j^{\rm th}$ pulse. The rate of pulses is $\Gamma$ and has units of s$^{-1}$; the pulses are assumed independent so the number in any time interval $\delta t$ is Poisson-distributed with mean $\Gamma \, \delta t$. Eq.~(1) of \citet{2013MNRAS.433.2275L} is similar except that the treatment here keeps the polarization index and the finite width of the pulse (i.e. it is taken as some function $\psi_j$ rather than a Dirac $\delta$-function). This makes the treatment here more general, but neither of these details is important to the final result.

In coming to the Earth, the signal passes through an ionized cloud of some dispersion measure DM, which introduces a phase delay given by $\phi(f)$. It follows that the received field is
\begin{equation}
\tilde E_\alpha(f) = \rme^{\rmi \phi(f)} \tilde E^{\rm em}_\alpha(f).
\label{eq:phdel}
\end{equation}
Written in the time domain, one has $E_\alpha = E^{\rm em}_\alpha\ast{\cal D}$, where
\begin{equation}
{\cal D}(t) = \int_{-\infty}^\infty \rme^{\rmi [\phi(f) - 2 \pi f t ] }\rmd f
\label{eq:D}
\end{equation}
is the delay kernel. (Plasma birefringence or Faraday rotation is ignored here as it is not germane to the problem at hand.)

An electric field observed with an actual detector $A$ is sensitive to some polarization state $p_{A\alpha}$ with some linear filter $\chi_A$:
\begin{equation}
\bar E_A(t) = \sum_\alpha p_{A\alpha} [\chi_A \ast E_\alpha](t).
\label{eq:Efilt}
\end{equation}
Here the bar is used to indicate an electric field processed through the polarization and spectral response of the telescope, feeds, and hardware and software filters, as opposed to the unbarred $E_\alpha(t)$ which denotes a free-space electric field incident on the telescope.
It is assumed that the bandpass filter falls off fast enough that we can approximate $\tilde\chi_A(f)=0$ for $f\le 0$ -- in particular, there is no direct-current (DC) sensitivity, i.e. that $\tilde\chi_A(0)=0$.

While Eq.~(\ref{eq:Efilt}) represents the electric field for an arbitrary bandpass, it is typical of a frequency channel in a radio receiver to have some central frequency $f_A$ and a bandpass shape (determined by some combination of hardware and software) with some width, $B_A$. In this case, the output electric field is determined by the central frequency and a bandpass function $\Delta_A$:
\begin{equation}
\bar E_A(t) = \rme^{-2\pi\rmi f_A t} \sum_\alpha p_{A\alpha} \int_{-\infty}^\infty E_\alpha(t') \rme^{2\pi\rmi f_A t'} \Delta_A(t-t')\,\rmd t'.
\label{eq:Efilt.1}
\end{equation}
The operation considered here is -- again following \citet{2013MNRAS.433.2275L} -- a multiplication against a local oscillator at frequency $f_A$ (i.e. the complex exponential $\rme^{2\pi\rmi f_A t}$) followed by a convolution with the bandpass-limiting filter $\Delta_A$, and finally by the prefactor that mixes the signal back to the original frequency.\footnote{The last step was done implicitly in \citet{2013MNRAS.433.2275L} in writing their Eq.~(2).} The filter $\Delta_A$ is taken to have temporal width $\sim B_A^{-1}$ so that it passes a frequency width of order $B_A$, and is normalized to unit transmission at the band center, $\tilde\Delta_A(0)=1$. A simple example is of course the top-hat in frequency $\Delta_A(\tau) = \sin (\pi B_A\tau)/(\pi\tau)$, though many other choices are possible. It is easily seen that Eq.~(\ref{eq:Efilt.1}) is equivalent to the general linear filter, Eq.~(\ref{eq:Efilt}), with
\begin{equation}
\chi_A(\tau) \equiv \rme^{-2\pi\rmi f_A\tau} \Delta_A(\tau)
~~\leftrightarrow~~
\tilde\chi_A(f) = \tilde\Delta_A(f-f_A).
\end{equation}

Note that since the filter $\chi$ is positive frequency only, $\chi_A(\tau)$ and hence $\bar E_A(t)$ are complex functions. The real and imaginary parts of $\bar E_A(t)$ correspond to the cosine-like and sine-like parts of $E_\alpha(t)$ (see Eq.~\ref{eq:Efilt.1}) in the same sense that a complex phasor is used to describe a real oscillatory function.

This paper is concerned with time-steady radio sources and, hence, we are interested in correlation functions of $\bar E_A(t)$. The ``1-point correlation function'' we write as $\langle \bar E_A(t) \rangle$, where $\langle ...\rangle$ denotes an ensemble average (which is equivalent for time-steady sources to a time average).  It vanishes trivially so long as the filter $A$ excludes the DC component, i.e. so long as $\tilde\chi_A(0)=0$. Thus, only 2-point and higher correlation functions are of interest.  The 2-point correlation function is 
\begin{equation}
C^E_{AB}(\delta t) \equiv \langle \bar E_A(t) \bar E_B^\ast(t+\delta t) \rangle,
\label{eq:CAB}
\end{equation}
where the average is taken over both the types of electron pulses (${\cal E}_j$ and $\psi_{j\alpha}$) and the pulse times ($t_j$) of each pulse.
In principle one may define the 2-point correlation function without the complex conjugate:
\begin{equation}
C^{(2,0)}_{AB}(\delta t) \equiv \langle \bar E_A(t) \bar E_B(t+\delta t)\rangle,
\label{eq:C20}
\end{equation}
but this is zero because $\bar E_A(t)$ and $\bar E_B(t+\delta t)$ are both positive-frequency functions (their Fourier transforms are zero at $f<0$). Hence, their product is also positive-frequency, and its integral over a sufficiently long period of time is zero. %Ergodicity then implies that the ensemble average is also zero.

Higher-order correlation functions may be defined analogously and some have a simple physical interpretation. For example, the correlation function of intensity fluctuations is a 4-point correlation function of the electric field, since the intensity is proportional to the square of the electric field.  We note that all the statistical properties of the radiation are encoded in $n$-point functions of the field.

The formalism presented here, and used in the main text, is based on classical signals and correlation functions rather than quantum states and operator expectation values. The use of classical electrodynamics in writing down Eq.~(\ref{eq:E}) is justified in detail in \S\ref{ss:QM} and Appendix~\ref{app:QM}.

\section{Effect of dispersion on the 2-point correlation function}
\label{sec:stat-c}

In \citet{2013MNRAS.433.2275L}, it was claimed that the 2-point correlation function of overlapping frequency channels with finite $|f_A-f_B|$ could be used to measure the dispersion measure to a time-steady synchrotron source. Using the tools of \S\ref{sec:stat-e}, we will now formally evaluate this correlation function and show that there is no DM information contained therein.  Section~\ref{sec:ex} discuses at length why the intuition that dispersion creates a temporal lag between the higher and lower-frequency channels fails.

Substituting our expression for $E_A(t)$ (Eq.~\ref{eq:Efilt}) into the expression for the correlation function (Eq.~\ref{eq:CAB}) yields
\begin{eqnarray}
C^E_{AB}(\delta t)
\!\!\!\! &=& \!\!\!\!
\sum_{\alpha\beta jk}
p_{A\alpha} p^\ast_{B\beta}
 \Big \langle
{\cal E}_j{\cal E}_k^\ast
[\chi_A\ast {\cal D}\ast \psi_{j\alpha}](t-t_j)
\nonumber \\ && \times~
[\chi_B^\ast\ast {\cal D}^\ast\ast \psi^\ast_{k\beta}](t+\delta t-t_k)
\Big \rangle.
\end{eqnarray}
One now splits the sum into two parts: one with $j=k$ and one with $j\neq k$. In each case we may replace the average over $t_j$ and summation over $j$ with $\Gamma\int_{-\infty}^\infty\,\rmd t_j$, giving
\begin{eqnarray}
C^E_{AB}(\delta t)
\!\!\!\! &=& \!\!\!\!
\Gamma \sum_{\alpha\beta} \int_{-\infty}^\infty\,\rmd t_j\,
p_{A\alpha} p^\ast_{B\beta}
\Big \langle
|{\cal E}_j|^2
[\chi_A\ast {\cal D}\ast \psi_{j\alpha}](t-t_j)
\nonumber \\ && \times~
[\chi_B^\ast\ast {\cal D}^\ast\ast \psi^\ast_{j\beta}](t+\delta t-t_j)
\Big \rangle
\nonumber \\ && +~ \Gamma^2
\sum_{\alpha\beta} \int_{-\infty}^\infty\,\rmd t_j\, \int_{-\infty}^\infty\,\rmd t_k\,
p_{A\alpha} p^\ast_{B\beta}
\nonumber \\ && \times~
\Big \langle
{\cal E}_j
{\cal E}_k^\ast
[\chi_A\ast {\cal D}\ast \psi_{j\alpha}](t-t_j)
\nonumber \\ && \times~
 [\chi_B^\ast\ast {\cal D}^\ast\ast \psi^\ast_{k\beta}](t+\delta t-t_k)
\Big \rangle.
\label{eq:CAB.1}
\end{eqnarray}
The second term (i.e. the term where the electric field comes from two different pulses) should be zero because distinct pulses do not correlate:  This term can be broken down into two separate integrals containing pulses $j$ and $k$, and each one individually vanishes upon integration over $t_j$ (or $t_k$) since $\tilde\chi_A(0)=\tilde\chi_B(0)=0$.  
Thus, only the first term of Eq.~(\ref{eq:CAB.1}) survives. With the replacement $\tau=t-t_j$, the integral over $t_j$ becomes an integral over $\tau$ such that
\begin{eqnarray}
C^E_{AB}(\delta t)
\!\!\!\! &=& \!\!\!\!
\Gamma \sum_{\alpha\beta} \int_{-\infty}^\infty\,\rmd \tau\,
p_{A\alpha} p^\ast_{B\beta}
 \Big \langle
|{\cal E}_j|^2
[\chi_A\ast {\cal D}\ast \psi_{j\alpha}](\tau)
\nonumber \\ && \times~
[\chi_B^\ast\ast {\cal D}^\ast\ast \psi^\ast_{j\beta}](\tau+\delta t)
\Big \rangle.
\label{eq:CAB.2}
\end{eqnarray}
Further simplification can be achieved by defining the time-reversal operator ${\cal R}$ via
\begin{equation}
[{\cal R}E](t) = E(-t)
~~\leftrightarrow~~
[\widetilde{{\cal R}E}](f) = \tilde E(-f).
\end{equation}
With the help of this operator and its trivial distributive property over convolution ${\cal R}(E\ast G) = ({\cal R}E)\ast({\cal R}G)$, Eq.~(\ref{eq:CAB.2}) reduces to
\begin{eqnarray}
C^E_{AB}(\delta t)
\!\!\!\! &=& \!\!\!\!
\Gamma \sum_{\alpha\beta} 
p_{A\alpha} p^\ast_{B\beta}
 \Big \langle
|{\cal E}_j|^2
[
({\cal R}\chi_A) \ast ({\cal{RD}}) \ast ({\cal R}\psi_{j\alpha})
\nonumber \\ &&
\ast~\chi_B^\ast\ast {\cal D}^\ast\ast \psi^\ast_{j\beta}](\delta t)
\Big \rangle.
\label{eq:CAB.3}
\end{eqnarray}

The final simplification involves the dispersion kernel ${\cal D}$. One sees that
\begin{equation}
[\widetilde{({\cal RD})\ast{\cal D}^\ast}](f) 
=[\widetilde{({\cal RD})}](f)[\widetilde{{\cal D}^\ast}](f)
=\tilde {\cal D}(-f) \tilde {\cal D}^\ast(-f)
= |\tilde{\cal D}(-f)|^2.
\end{equation}
However, the dispersion kernel has a Fourier transform $\tilde{\cal D}(f) = \rme^{\rmi\phi(f)}$, where $\phi(f)$ is a real function. It follows that $|\tilde{\cal D}(-f)|^2=1$ and so the ${\cal{RD}}$ and ${\cal D}^\ast$ in Eq.~(\ref{eq:CAB.3}) cancel out:
\begin{equation}
C^E_{AB}(\delta t)
=
\Gamma \sum_{\alpha\beta} 
p_{A\alpha} p^\ast_{B\beta}
 \Big \langle
|{\cal E}_j|^2
[
({\cal R}\chi_A) \ast ({\cal R}\psi_{j\alpha})
\ast\chi_B^\ast\ast \psi^\ast_{j\beta}](\delta t)
\Big \rangle.
\label{eq:CAB.4}
\end{equation}
The correlation function of $E_A(t)$ with $E_B(t)$ depends on the strength and shape of the pulses as well as the polarization and bandpass response of the detectors. However, the dispersion dependence has dropped out. Therefore, {\em dispersion has no effect on the correlation function of observables that are linear in the electric field}.

\section{A worked example: correlation of overlapping frequency channels}
\label{sec:ex}

It is instructive to explicitly work through the electric field correlation function in a simple example to see how it does not depend on DM.  The example considered here follows that in \citet{2013MNRAS.433.2275L}, taking $\psi_{j\alpha}(t)$ to be a $\delta$-function and considering a single polarization (so that the polarization indices need not be kept). The bandpass filter shapes $\tilde\Delta_A(f)$ will be taken to be identical for all channels, and only their central frequencies will differ.  The bandpass function $\Delta(f)$ will be kept arbitrary at first, and then two explicit examples will be given: a Gaussian bandpass and a tophat bandpass.

We approximate the phase delay (cf. Eq. \ref{eq:phdel}) as quadratic in frequency and given by
\begin{equation}
\phi(f) = \phi_A + 2\pi T_A (f-f_A) + \pi D(f-f_A)^2.
\label{eq:lin-disp}
\end{equation}
(We approximate $\phi$ to be a quadratic function of $f$ across all channels so that $D$ does not require a subscript.)
In accordance with Eq.~(\ref{eq:D}), the stationary-phase time delay as a function of frequency is $T(f) = (2\pi)^{-1} \rmd\phi(f)/\rmd f$, so that $T_A=T(f_A)$ is the delay at the center of the frequency band and $D = \rmd T/\rmd f$ encapsulates how the pulses is broadened by dispersion (units: s GHz$^{-1}$).  Note that for plasma dispersion $T>0$ but $D<0$. In \S\ref{sec:int-fluct}, we will need to use some numerical estimates for $D$; its relation to the usual dispersion measure is
%T = 4 ({\rm DM}/10^3\,{\rm pc\,cm}^{-3}) f_{\rm GHz, A}^{-2}~{\rm s},
\begin{equation}
D = -8 \left( \frac{\rm DM}{10^3\,{\rm pc\,cm}^{-3}}\right) f_{\rm GHz}^{-3} ~{\rm s ~GHz^{-1}},
\label{eq:typdisp}
\end{equation}
where $10^3$ pc$\,$cm$^{-3}$ is roughly the expected DM  to a source at $z=1$.
%so typical dispersions for cosmological sources would be of order a few s$\,$GHz$^{-1}$.

\subsection{General formula for arbitrary bandpass}

The dispersion kernel is the Fourier transform of $e^{i \phi(f)}$ or
\begin{eqnarray}
{\cal D}(t) \!\!\!\! &=& \!\!\!\!
\rme^{\rmi\phi_A} \int_{-\infty}^\infty \rme^{2\pi\rmi [T_A(f-f_A) + D(f-f_A)^2/2 -ft]} \,\rmd f
\nonumber \\ &=& \!\!\!\!
\rme^{\rmi\phi_A}\rme^{-2\pi\rmi f_At} \int_{-\infty}^\infty \rme^{2\pi\rmi [(T_A-t)(f-f_A) + D(f-f_A)^2/2]} \,\rmd f
\nonumber \\ &=& \!\!\!\!
\rme^{\rmi\phi_A}\rme^{-2\pi\rmi f_At} \rme^{-\pi \rmi (t-T_A)^2/D} \int_{-\infty}^\infty \rme^{\pi\rmi D(f-f_0)^2} \,\rmd f
\nonumber \\ &=& \!\!\!\!
\frac1{\sqrt{-\pi D}} \rme^{\rmi(\phi_A-\pi/4)}\rme^{-2\pi\rmi f_At} \rme^{-\pi \rmi (t-T_A)^2/D},
\end{eqnarray}
where we have set $f_0 = f_A + (t-T_A)/D$ in the third line.

With this ${\cal D}(t)$, ignoring polarization, and if the emitted field is a sequence of $\delta$-functions, the received field is
\begin{equation}
E(t) = \sum_j \frac{{\cal E}_j}{\sqrt{-\pi D}} \rme^{\rmi(\phi_A-\pi/4)}\rme^{-2\pi\rmi f_A(t-t_j)} \rme^{-\pi \rmi (t-t_j-T_A)^2/D}.
\end{equation}
The filtered field is then given by Eq.~(\ref{eq:Efilt.1}):
\begin{eqnarray}
\bar E_A(t)
\!\!\!\! &=& \!\!\!\!
\sum_j \frac{{\cal E}_j \rme^{\rmi(\phi_A-\pi/4)}}{\sqrt{-\pi D}} \rme^{-2\pi\rmi f_A(t-t_j)}
\int_{-\infty}^\infty \rme^{-\pi \rmi (t'-t_j-T_A)^2/D} \Delta(t-t')\,\rmd t'
\nonumber\\
\!\!\!\! &=& \!\!\!\!
\sum_j {\cal E}_j \rme^{\rmi \phi_A}
\rme^{-2\pi\rmi f_A(t-t_j)} \Delta^{(D)}(t-t_j-T_A),
\label{eq:ex.1}
\end{eqnarray}
where 
\begin{equation}
\Delta^{(D)}(\tau) \equiv \frac{\rme^{-\rmi\pi/4}}{\sqrt{-\pi D}} \int_{-\infty}^\infty \rme^{-\pi\rmi \tau'{}^2/D} \Delta(\tau-\tau')\,\rmd\tau'
\label{eq:disp}
\end{equation}
is the dispersed bandpass function.
The dispersed bandpass is formally equivalent to taking the original bandpass and smearing it with a Gaussian of complex width $\sqrt{D/(2\pi\rmi)}$; as a check, one may verify that for $D\rightarrow 0$ the original and dispersed bandpasses are equal, $\Delta^{(0)}(\tau)=\Delta(\tau)$. One expects the effect of dispersion on the signal within band to be modest when the smearing width is less than the intrinsic width, $|D|^{1/2}<B^{-1}$: this is the standard smearing criterion, encapsulated in \citet[][Eq.~10]{2013MNRAS.433.2275L}. Indeed, the fractional effect on the variance of $\Delta^{(D)}$ should be of order $|D|B^2$.

Using the notation of Eq.~(\ref{eq:ex.1}), the correlation function of $\bar E_A(t)$ with $\bar E_B(t)$ is
\begin{eqnarray}
C^E_{AB}(\delta t)
\!\!\!\! &=& \!\!\!\!
\Gamma \langle |{\cal E}_j|^2 \rangle
\,\rme^{\rmi(\phi_A-\phi_B)}
\int_{-\infty}^\infty
\rme^{-2\pi\rmi f_A(t-t_j)} \Delta^{(D)}(t-t_j-T_A)
\nonumber \\ && \times
\rme^{2\pi\rmi f_B(t+\delta t-t_j)} \Delta^{(D)\ast}(t+\delta t-t_j-T_B)
\,\rmd t_j.
\end{eqnarray}
Setting $T_{BA} = T_B-T_A$ and $f_{BA} = f_B-f_A$, and defining $\bar f = (f_A+f_B)/2$ and $\bar T = (T_A+T_B)/2$, we may make the substitution $\tau = t - t_j - \bar T + \delta t/2$. Also we recognize that $\phi_A-\phi_B = -2\pi\bar Tf_{BA}$. With these simplifications,
\begin{eqnarray}
C^E_{AB}(\delta t)
\!\!\!\! &=& \!\!\!\!
\Gamma \langle |{\cal E}_j|^2 \rangle\,
\rme^{2\pi \rmi \bar f \delta t}
\int_{-\infty}^\infty
\rme^{2\pi\rmi f_{BA}\tau}
\Delta^{(D)}\left(\tau + \frac{T_{BA}-\delta t}2\right)
\nonumber \\ && \times
\Delta^{(D)\ast}\left(\tau - \frac{T_{BA}-\delta t}2\right)
\,\rmd \tau.
\label{eq:ex2}
\end{eqnarray}
In Eq.~(\ref{eq:ex2}), the prefactor consists only of an overall normalization and a phase ($2\pi\bar f\delta t$) that is a property only of the instrument and software; thus only the integral is interesting, which is given by
\begin{equation}
{\cal J}(s) = \int_{-\infty}^\infty
\rme^{2\pi\rmi f_{BA}\tau}
\Delta^{(D)}(\tau + s)
\Delta^{(D)\ast}(\tau - s)
\,\rmd \tau,
\label{eq:ex3I}
\end{equation}
so that
\begin{equation}
C^E_{AB}(\delta t)
= \Gamma \langle |{\cal E}_j|^2 \rangle\,
\rme^{2\pi \rmi \bar f \delta t} {\cal J}\left( \frac{T_{BA}-\delta t}2 \right).
\label{eq:ex3C}
\end{equation}

\subsection{Some comments on the structure of the correlation integral}

One might expect the modulus of the correlation function to be maximal when $\delta t\approx T_{BA}$ as this likely maximizes the overlap of the functions inside the integral in Eq.~(\ref{eq:ex3I}). Indeed, if $\Delta^{(D)}(\tau)$ were a time-symmetric function such as a Gaussian of real standard deviation or a sinc-function, then $|{\cal J}(s)|$ would be symmetric around $s=0$, and the correlation function would peak exactly at $\delta t = T_{BA}$. In this way, it would be possible to measure $T_{BA}$ and hence $D \approx -2 T_{BA}/f_{BA}$ from the correlation of two adjacent frequency channels.

However, since the time delay between the two frequency channels is $Df_{BA}$, and we must have $f_{BA}\lesssim B$ in order for the signals at the two frequencies to be coherent over a correlation time $\sim B^{-1}$, the ratio of the delay to the correlation time is $|Df_{BA}| B \lesssim |D|B^2$. Thus according to the preceding discussion (following Eq.~\ref{eq:disp}), the time delay between the frequencies $f_A$ and $f_B$ is of the same order of magnitude in terms of fractional effect on ${\cal J}(s)$ as the deviation of the dispersed response $\Delta^{(D)}$ from the instrumental response $\Delta$. One must determine whether $|{\cal J}(s)|$ is really peaked at, or symmetric around, $s=0$.

It is trivial that the integrand in Eq.~(\ref{eq:ex3I}) has a modulus that is symmetric under $s\leftrightarrow-s$, so our attention turns instead to the phase structure.
Of particular interest is the possibility that $\Delta^{(D)}(t)$ could exhibit a ``phase acceleration,'' i.e. that $\alpha = \rmd^2[\arg \Delta^{(D)}(t)]/\rmd t^2$ may be nonzero. In this case, and taking for example $f_{BA}>0$ and $\alpha>0$, the argument of the integrand in Eq.~(\ref{eq:ex3I}) should vary more slowly for $s<0$ than for $s>0$, since
\begin{eqnarray}
\!\!\!\! && \!\!\!\!
\frac\rmd{\rmd\tau}\arg [
\rme^{2\pi\rmi f_{BA}\tau}
\Delta^{(D)}(\tau + s)
\Delta^{(D)\ast}(\tau - s)
]
\nonumber \\
&& = 
2\pi f_{BA} + \frac{\rmd \arg\Delta^{(D)}(\tau+s)}{\rmd\tau}
- \frac{\rmd \arg\Delta^{(D)}(\tau-s)}{\rmd\tau}
\nonumber \\
&&\sim 2\pi f_{BA} + 2\alpha s.
\end{eqnarray}
(The last step is only schematic since -- except in special cases such as the Gaussian bandpass -- $\alpha$ is not constant.)
Thus if $\alpha>0$, $|I(s)|$ should be enhanced for $s<0$ and suppressed for $s>0$ (and the reverse if $\alpha<0$). This asymmetry should manifest itself in the observed correlation function $C_{AB}(\delta t)$.

This situation is clarified next for two explicit cases: a Gaussian and a tophat. In each case, $\Delta^{(D)}(t)$ will be evaluated, and it will be shown that it exhibits a positive phase acceleration.

\subsection{Case of Gaussian bandpass}
\label{ss:G}

\begin{figure*}
\includegraphics[width=6.5in]{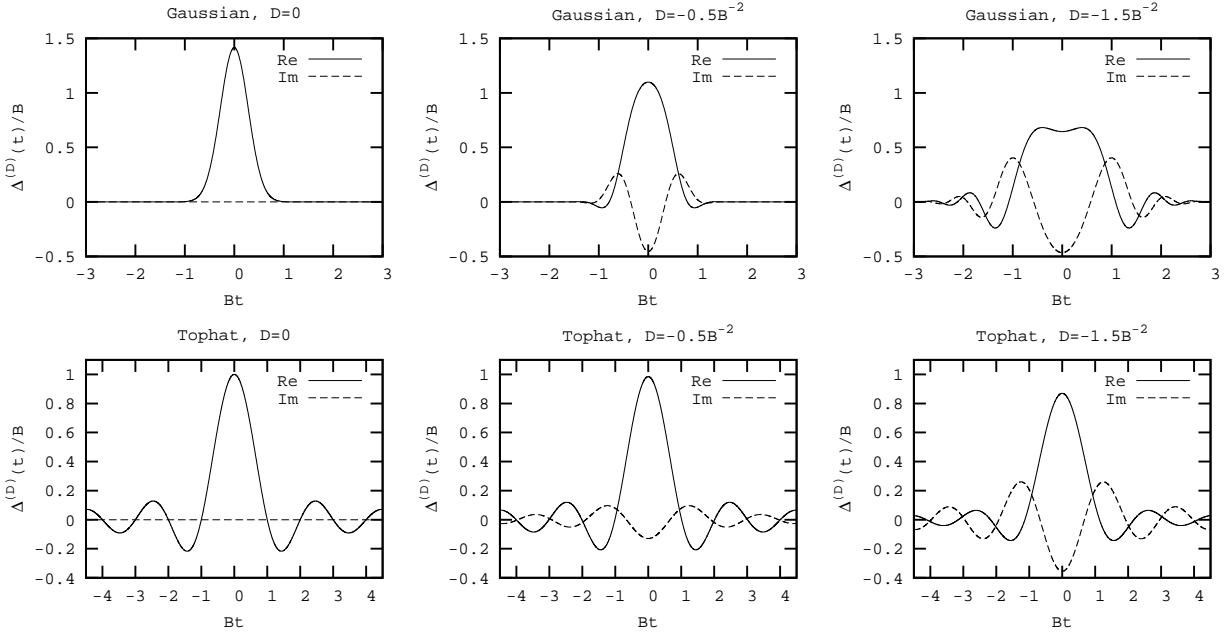}
\caption{\label{fig:deltadispersed}The real and imaginary parts of the dispersed bandpass function, $\Delta^{(D)}(t)$, for $D=0$, $-0.5B^{-2}$, and $-1.5B^{-2}$, for the Gaussian (upper row) and tophat (lower row) bandpasses. The scales are different for the two rows. Note the behaviour of the phase in the dispersed cases: $\Delta^{(D)}(t)$ is positive-frequency at $t<0$ (the imaginary part leads the real part) and negative-frequency at $t>0$ (the real part leads the imaginary part). In the integral for ${\cal J}(s)$, Eq.~(\ref{eq:ex3I}), this phase acceleration leads to a peak in the correlation integral at positive $s$ (for $f_B>f_A$ or $f_{BA}>0$), in accordance with the principle of stationary phase. It is this offset in the peak of the correlation integral between pulses in frequency channels $f_A$ and $f_B$ that cancels the dispersion-induced delay in arrival time at frequency $f_A$ relative to $f_B$, and leads to zero observed delay in the peak of the cross-correlation function between the two channels.}
\end{figure*}

The Gaussian bandpass is the simplest choice for the purposes of analytic calculation. It is defined by
\begin{equation}
\tilde\Delta(f) = \rme^{-f^2/2\sigma_f^2}
~~\leftrightarrow~~
\Delta(t) = \frac1{\sqrt{2\pi}\,\sigma_t} \rme^{-t^2/2\sigma_t^2},
\end{equation}
where $\sigma_f = 1/(2\pi\sigma_t)$. The power-equivalent bandwidth\footnote{Other definitions of bandwidth are possible and have no effect on the calculation; this one is chosen for definiteness, and for consistency with the bandwidth $B$ in \citet{2013MNRAS.433.2275L}.} is
\begin{equation}
B \equiv \int_{-\infty}^\infty |\tilde\Delta(f)|^2\,\rmd f = \sqrt{\pi}\,\sigma_f = \frac1{2\sqrt{\pi}\,\sigma_t}.
\label{eq:PEB}
\end{equation}
%\flagthis{[May want to use different symbol as B has two meanings here]}
According to Eq.~(\ref{eq:disp}), the dispersed bandpass is obtained by convolving the instrumental response $\Delta(t)$ with a Gaussian of variance $D/(2\pi\rmi)$. Thus one has
\begin{equation}
\Delta^{(D)}(t) = \frac1{\sqrt{2\pi}\,\sigma_t^{(D)}} \rme^{-t^2/2\sigma_t^{(D)\,2}},
\end{equation}
where
\begin{equation}
\sigma_t^{(D)} \equiv \sqrt{ \sigma_t^2 + \frac{D}{2\pi\rmi}}.
\end{equation}
Note that since $D<0$, $D/(2\pi\rmi)$ is on the positive imaginary axis, so that $\sigma_t^{(D)}$ lies in the first octant, i.e. $0<\arg \sigma_t^{(D)}<\pi/4$. Furthermore, $\sigma_t^{(D)\,-2}$ lies in the fourth quadrant, so that $\Delta^{(D)}(t)$ takes the form of a Gaussian envelope with varying phase. Decomposing $\sigma_t^{(D)\,-2} = a-\rmi b$ with $a$ and $b$ real gives
\begin{equation}
a = \frac{\sigma_t^2}{\sigma_t^4+(D/2\pi)^2}
~~{\rm and}~~
b = \frac{-D/2\pi}{\sigma_t^4+(D/2\pi)^2}.
\end{equation}
The phase of $\Delta^{(D)}(t)$ varies as $\arg\Delta^{(D)}(t) = {\rm constant} + bt^2/2$. It is this ``phase acceleration'' of $\Delta^{(D)}(t)$ that will lead to a shift in the peak of $|{\cal J}(s)|$. One notes that the phase shift at $t=0$ is negative, but that the second derivative is positive. To lowest order in $D$, we have $\rmd^2[\arg\Delta^{(D)}(t)]/\rmd t^2|_{t=0} = -8\pi DB^4$. The dispersed bandpass, $\Delta^{(D)}(t)$, for this Gaussian case are plotted in the top panels of Fig.~\ref{fig:deltadispersed}.   The top-left panel shows $D=0$, the top-middle shows $D=-0.5B^{-2}$, and the top-right shows $D=-1.5B^{-2}$.  The phase acceleration is clearly visible for finite $D$: $\Delta^{(D)}(t)$ is positive-frequency at $t<0$ (the imaginary part leads the real part) and negative-frequency at $t>0$ (the real part leads the imaginary part).

Returning to Eq.~(\ref{eq:ex3I}), one finds 
\begin{eqnarray}
{\cal J}(s) \!\!\!\! &=& \!\!\!\! \frac1{2\pi|\sigma_t^{(D)}|^2}
\int_{-\infty}^\infty
\rme^{2\pi\rmi f_{BA}\tau}
\rme^{-(\tau+s)^2/2\sigma_t^{(D)\,2}}
\rme^{-(\tau-s)^2/2\sigma_t^{(D)\ast\,2}}
\,\rmd \tau
\nonumber \\
&=& \!\!\!\!
\frac{\rme^{-as^2}}{2\pi|\sigma_t^{(D)}|^2}
\int_{-\infty}^\infty
\rme^{2\rmi(\pi f_{BA} + bs)\tau}
\rme^{-a\tau^2}
\,\rmd \tau
\nonumber \\
&=& \!\!\!\!
\frac{\rme^{-as^2}}{2\pi^{1/2} a^{1/2}|\sigma_t^{(D)}|^2}
\rme^{-(\pi f_{BA}+bs)^2/a}
\nonumber \\
&=& \!\!\!\!
\frac{1}{2\sqrt\pi\, a^{1/2}|\sigma_t^{(D)}|^2}
\rme^{-[(a^2+b^2)s^2 + 2\pi f_{BA}bs + \pi^2 f_{BA}^2]/a}
\nonumber \\
&=& \!\!\!\!
\frac1{2\sqrt\pi\,\sigma_t}
\rme^{-\pi f_{AB}^2\sigma_t^2} \rme^{-(s - Df_{BA}/2)^2/\sigma_t^2}.
\label{eq:answer-gauss}
\end{eqnarray}
Here the first equality is substitution into Eq.~(\ref{eq:ex3I}); the second is an algebraic conversion of $\sigma_t^{(D)}$ into $a$ and $b$; the third is a Gaussian integral; the fourth is an expansion of the exponent in terms of $s$; and the fifth is a simplification using the identities $a^{1/2}|\sigma_t^{(D)}|^2 = \sigma_t$ (for the prefactor), and $a/(a^2+b^2) = \sigma_t^{2}$ and $b/(a^2+b^2) = -D/(2\pi)$ (for completing the square in the exponent).

Equation~(\ref{eq:answer-gauss}) implies that the overlap integral between the two dispersed response functions is not at $s=0$ but instead at $s=-Df_{BA}/2$. This is a direct consequence of the phase acceleration term $b\neq 0$. One then concludes that the observed correlation function $C_{AB}(\delta t)$ peaks not at $\delta t = T_{BA}$ but at
\begin{equation}
\delta t = T_{BA} - 2s_{\rm peak} = T_{BA} - D f_{BA} = 0.
\end{equation}
The last equality is the first step in this section (\S\ref{sec:ex}) where the fact that $D=T_{BA}/f_{BA}$ has been explicitly used.
Thus one concludes that even though the signal at frequency $f_B$ is delayed relative to that at $f_A$, the warping of the signal within each band due to dispersion produces an equal and opposite shift of the peak of the correlation function, leading to no net observable effect.

\subsection{Case of tophat bandpass}
\label{ss:T}

The tophat bandpass is
\begin{equation}
\Delta(t) = \frac{\sin(\pi Bt)}{\pi t}
~~\leftrightarrow~~
\tilde\Delta(f) = \Pi\left( \frac fB \right),
\end{equation}
where the unit tophat function is $\Pi(x) = 1$ if $|x|<\frac12$ and 0 otherwise. It is used as an example in \citet{2013MNRAS.433.2275L}. The calculation for this bandpass is more involved than for the Gaussian, but a similar result will be derived. The power-equivalent bandpass (see Eq.~\ref{eq:PEB}) is trivially shown to be $B$.

The dispersed bandpass is given by
\begin{equation}
\Delta^{(D)}(t) = \frac{\rme^{-\rmi \pi/4}}{\sqrt{-\pi D}} \int_{-\infty}^\infty
\rme^{-\pi \rmi t'{}^2/D} \frac{\sin[\pi B(t-t')]}{\pi (t-t')} \,\rmd t'.
\label{eq:dd}
\end{equation}
The simplest form can be obtained by substituting the relation
\begin{equation}
\frac{\sin [\pi B(t-t')]}{\pi (t-t')} = \frac1{2\pi } \int_{-\pi B}^{\pi B} \rme^{\rmi u(t-t')}\,\rmd u
\end{equation}
into Eq.~(\ref{eq:dd}), and then performing the Gaussian $t'$ integral to get:
\begin{equation}
\Delta^{(D)}(t) = \frac{1}{2\pi} \int_{-\pi B}^{\pi B}
\rme^{\rmi [D u^2/(4\pi) + ut]}
\,\rmd u.
\label{eq:ddt}
\end{equation}
The substitution $u' = u - 2\pi t/D$ turns this into a Fresnel integral:
\begin{eqnarray}
\Delta^{(D)}(t)
\!\!\!\! &=& \!\!\!\!
\frac{1}{2\pi } \rme^{-\rmi\pi t^2/D} \int_{-\pi B - 2\pi t/D}^{\pi B - 2\pi t/D}
\rme^{\rmi D u'{}^2/(4\pi)}
\,\rmd u'
\nonumber \\
\!\!\!\! &=& \!\!\!\!
\frac{\rme^{\rmi\pi t^2/(2w^2)}}{2w} \left[
{\cal F}^\ast\left(wB+\frac tw\right)
-{\cal F}^\ast\left(-wB+\frac tw\right)
\right],
\label{eq:ddf}
\end{eqnarray}
where $w\equiv \sqrt{-D/2}$ has units of time and ${\cal F}$ is the Fresnel integral\footnote{This is given by ${\cal F}(z) = C(z) + \rmi S(z)$ in the notation of \citet[][\S7.3]{1972hmfw.book.....A}.},
\begin{equation}
{\cal F}(z) = \int_0^z \rme^{\rmi\pi \varsigma^2/2}\,\rmd\varsigma.
\end{equation}
The Fresnel integral as a function of real $z$ traces out the familiar ``Cornu spiral'' in the complex plane, arcing from ${\cal F}(-\infty)=-\frac12-\frac12\rmi$ to ${\cal F}(\infty)=\frac12+\frac12\rmi$. According to Eq.~(\ref{eq:ddf}), the bandpass function $\Delta^{(D)}(t)$ is the separation vector between two points on the spiral with parameter $z=t/w\pm wB$, with the instantaneous direction of motion at $z=t/w$ removed by a phase rotation $\rme^{\rmi\pi t^2/(2w^2)}$, and with the normalizing factor $2w$. The dispersed tophat bandpass function is shown in the bottom panels of Fig.~\ref{fig:deltadispersed}.

Equation~(\ref{eq:ddf}) shows that there is a phase acceleration of $\Delta^{(D)}(t)$ analogous to that which occurred for the Gaussian case. There is no analytic expression for the phase in this case, but it is possible to do a Taylor expansion\footnote{This is straightforward by brute force expansion of Eq.~(\ref{eq:ddt}).} of $\arg\Delta^{(D)}(t)$ to order $w^2t^2$ or $Dt^2$ (in the two parameters $w$ and $t$) and thus obtain the phase acceleration at the center of the pulse:
\begin{equation}
\arg \Delta^{(D)}(t) = \frac\pi{12}DB^2 - \frac{\pi^3}{90} DB^4 t^2 + ...,
\end{equation}
i.e. there is a negative phase shift at $t=0$ but with an instantaneous phase acceleration of $-(\pi^3/45)DB^4$. This is just as for the Gaussian case, albeit with a different prefactor.
%
%\begin{eqnarray}
%\arg \Delta^{(D)}(t)
%\!\!\!\! &=& \!\!\!\! \frac{\pi t^2}{2w^2} - \arg \int_{t/w-wB}^{t/w+wB} \rme^{\rmi\pi \varsigma^2/2}\,\rmd\varsigma
%\nonumber \\
%\!\!\!\! &=& \!\!\!\! - \arg \int_{-1}^{1} \rme^{\rmi\pi [(t/w+wBu)^2-(t/w)^2]/2}\,\rmd u
%\nonumber \\
%\!\!\!\! &=& \!\!\!\! - \arg \int_{-1}^{1} \rme^{\rmi\pi [ Btu + (wBu)^2/2 ]} \,\rmd u
%\nonumber \\
%\!\!\!\! &=& \!\!\!\! - \arg \int_{-1}^{1} \Bigl(
%1 - \frac14\rmi\pi D B^2 u^2
%- \frac12\pi^2B^2t^2u^2
%\nonumber \\ &&
%+ \frac18\rmi\pi^3 D B^4t^2u^4 + ...
%\Bigr) \,\rmd u
%\nonumber \\
%\!\!\!\! &=& \!\!\!\! - \arg \Bigl(
%1 - \frac1{12}\rmi\pi D B^2
%- \frac16\pi^2B^2t^2
%+ \frac1{40}\rmi\pi^3 D B^4t^2 + ...
%\Bigr) 
%\nonumber \\
%\!\!\!\! &=& \!\!\!\!
%\frac\pi{12}DB^2 - \frac{\pi^3}{90} DB^4 t^2 + ...,
%\end{eqnarray}
%where in the fourth equality terms in the expansion with odd order in $u$ have been dropped since they integrate to zero. There is a negative phase shift at $t=0$ but with an instantaneous phase acceleration of $-(\pi^3/45)DB^4$. This is just as for the Gaussian case, albeit with a different prefactor.

One is now interested in the integral ${\cal J}(s)$, which is obtained by integrating two copies of $\Delta^{(D)}(t)$. Substituting Eq.~(\ref{eq:ddf}) into Eq.~(\ref{eq:ex3I}) gives
\begin{eqnarray}
{\cal J}(s)\!\!
 \!\!\!\! &=& \!\!\!\!
\int_{-\infty}^\infty
\frac{\rme^{2\pi\rmi (f_{BA} + s/w^2)\tau}}{4w^2}
 \left[
{\cal F}^\ast\left(wB+\frac {\tau+s}w\right)
-{\cal F}^\ast\left(-wB+\frac {\tau+s}w\right)
\right]
\nonumber \\ && \times
 \left[
{\cal F}\left(wB+\frac {\tau-s}w\right)
-{\cal F}\left(-wB+\frac {\tau-s}w\right)
\right]
\rmd \tau
\nonumber \\
\!\!\!\! &=& \!\!\!\!
\int_{-\infty}^\infty
\frac{\rme^{2\pi\rmi (f_{BA} + s/w^2)\tau}}{4w^2}
\int_{-wB}^{wB}
\int_{-wB}^{wB}
\rme^{-\rmi\pi [(\tau+s)/w + x]^2/2}
\nonumber \\ && \times
\rme^{\rmi\pi [(\tau-s)/w + y]^2/2}
\,\rmd x\,\rmd y\,\rmd\tau
\nonumber \\
\!\!\!\! &=& \!\!\!\!
\int_{-\infty}^\infty
\frac{\rme^{2\pi\rmi (f_{BA} + s/w^2)\tau}}{4w^2}
\int_{-wB}^{wB}
\int_{-wB}^{wB}
\rme^{\rmi (y^2-x^2)/2}
\nonumber \\ && \times
\rme^{\rmi\pi [ - 2\tau s/w^2 + \tau (y-x)/w - s(x+y)/w]}
\,\rmd x\,\rmd y\,\rmd\tau,
\end{eqnarray}
where in the second line the Fresnel integral has been re-expanded using the fundamental theorem of calculus. The integral appearing in the last expression can be simplified by performing the $\tau$ integral first, which leads to a $\delta$-function:
\begin{equation}
{\cal J}(s) =
\frac\pi{2w^2}\!
\int_{-wB}^{wB}\!
\int_{-wB}^{wB}\!
\rme^{\rmi\pi [(y^2-x^2)/2 - s(x+y)/w]}
\delta\left(
2\pi f_{BA} + \pi\frac{y-x}w
\right)
\,\rmd x\,\rmd y.
\label{eq:temp-2111}
\end{equation}
The $\delta$-function enforces that $y-x = 2wf_{BA}$, and hence that $(y^2-x^2)/2=wf_{BA}(y+x)$. Switching coordinates to $z=(x+y)/2$ and $v=(y-x)/w$, so that $\rmd x\,\rmd y = w\, \rmd z \,\rmd v$, and then trivially integrating the $\delta$-function, one finds
\begin{equation}
{\cal J}(s) =
\frac1{2w}
\int_{-z_{\rm max}}^{z_{\rm max}}
\rme^{2\pi \rmi z (wf_{BA} - s/w)}
\,\rmd z,
\label{eq:temp-2118}
\end{equation}
where the range of integration is such that $x=z -wf_{BA}$ and $y=z+wf_{BA}$ are both between $-wB$ and $wB$ -- i.e. we have $z_{\rm max} = w(B-|f_{BA}|)$ if $|f_{BA}|<B$ and 0 otherwise. The integral is then
\begin{equation}
{\cal J}(s) = \frac{\sin [2\pi (z_{\rm max}/w)(w^2f_{BA}-s)]}{ 2\pi(w^2f_{BA}-s)}.
\end{equation}
Substituting back in the expressions for $w$ and $z_{\rm max}$:
\begin{equation}
{\cal J}(s) = \Theta(B-|f_{BA}|) \frac{\sin [2\pi (B-|f_{BA}|)(s+Df_{BA}/2)]}{ 2\pi(s+Df_{BA}/2) }.
\label{eq:istophat}
\end{equation}

From Eq.~(\ref{eq:istophat}), one sees that the overlap function again depends only on the dispersion $D$ through an overall offset of the horizontal scale: ${\cal J}(s)$ is shifted to be centered at $s=-Df_{BA}/2$. This is the same behaviour as was found in \S\ref{ss:G} for the Gaussian integral, and has the same consequence: that the phase acceleration of $\Delta^{(D)}(t)$ introduces an offset in the correlation of the two channels that exactly cancels the delay difference $T_{BA}$ between the two central frequencies.

\section{Intensity fluctuations}
\label{sec:int-fluct}

The intensity fluctuations from a source are related to the $4$-point correlation function of the electric field. Defining the mean intensity in a channel $\bar I_A = \langle |\bar E_A(t)|^2 \rangle$ and an intensity fluctuation
\begin{equation}
\delta I_A(t) = |\bar E_A(t)|^2 - \bar I_A,
\end{equation}
one can find the intensity correlation function in two channels:
\begin{equation}
C^{\delta I}_{AB}(\delta t) = \langle \bar E_A(t) \bar E_A^\ast(t) \bar E_B(t + \delta t) \bar E_B^\ast(t+ \delta t) \rangle - \bar I_A \bar I_B.
\label{eqn:CIAB}
\end{equation}
The next task is to evaluate the intensity correlation function and determine how it depends on dispersion measure. It will be shown that there is indeed a dependence on DM: the argument leading to Eq.~(\ref{eq:CAB.4}) that showed that the 2-point correlation function was independent of DM does not apply to higher-order correlation functions, since only in the 2-point case can the integral over pulse epochs $t_j$ be converted to a convolution of ${\cal{RD}}$ and ${\cal D}^\ast$. (For higher-point correlation functions, a more complicated set of integrals over ${\cal D}$ applies.) However only the {\em connected part}\footnote{The connected part of the 4-point function of any set of zero-mean variables is defined as $\langle wxyz\rangle_{\rm c} \equiv \langle wxyz \rangle - \langle wx\rangle\langle yz\rangle - \langle wy\rangle\langle xz\rangle - \langle wz\rangle\langle xy\rangle$. It vanishes for Gaussian fields.} of the electric field 4-point function can depend on DM, since the disconnected part consists of 2-point functions and is thus independent of DM as shown in \S\ref{sec:stat-c}. If the number of independent emitting electrons is large, we will find that the disconnected part dominates. Section \S\ref{ss:detectability} presents an order-of-magnitude evaluation of the importance of the connected terms that depend on DM: for parameters appropriate to a realistic AGN they are immeasurably small.  Finally, we discuss the origin of this counterintuitive result in \S\ref{ss:lieuduan} and discuss why our result differs from the conclusion in \citet{2013ApJ...763L..44L}. 

\subsection{Computation of the intensity correlation function}
\label{ss:ICF}

Substituting into Eq.~(\ref{eqn:CIAB}) yields
\begin{eqnarray}
 C^{\delta I}_{AB}(\delta t)& = & \sum_{jklm} \Big \langle
{\cal E}_j {\cal E}_k^\ast {\cal E}_l {\cal E}_m^\ast
[{\cal D} \ast \Psi_{jA}](t-t_j)
~[{\cal D}^\ast\ast\Psi_{kA}^\ast](t-t_k)
\nonumber \\ && \times
[{\cal D} \ast \Psi_{lB}](t+\delta t-t_l)
~[{\cal D}^\ast\ast\Psi_{mB}^\ast](t+\delta t-t_m)
 \Big \rangle
\nonumber \\ &&
- \sum_{jklm} \Big \langle
\epsilon_j \epsilon_k^\ast
[{\cal D} \ast \Psi_{jA}](t-t_j)
~[{\cal D}^\ast\ast\Psi_{kA}^\ast](t-t_k) \Big \rangle
\nonumber \\ && \times \Big \langle
\epsilon_l \epsilon_m^\ast
[{\cal D} \ast \Psi_{lB}](t-t_l)
~[{\cal D}^\ast\ast\Psi_{mB}^\ast](t-t_m)
 \Big \rangle.
\label{eq:CIx}
\end{eqnarray}
To reduce clutter, we have introduced the notation
\begin{equation}
\Psi_{jA}(\tau) \equiv \sum_\alpha p_{A\alpha} [\chi_A \ast \psi_{j\alpha}](\tau),
\end{equation}
which is the electron pulse ($\psi_{j\alpha}$) observed through the instrument bandpass $\chi_A$ and polarization state $p_{A\alpha}$. As always, it is assumed that there is no DC response: $\tilde \chi_A(0)=0$, or $\int_{-\infty}^\infty \Psi_{jA}(\tau)\,\rmd\tau = \tilde\Psi_{jA}(0) = 0$.

The nonzero contributions to $C^{\delta I}_{AB}$ result from each index being equal to at least one other index (on account of the ``no DC response'' condition). Breaking the first term of Eq.~(\ref{eq:CIx}) into these different components yields several terms: (i) a term with $j=k$ and $l=m$; (ii) a term with $j=l$ and $k=m$; (iii) a term with $j=m$ and $k=l$; and (iv) a term with $j=k=l=m$. The term of the form (i) cancels the second term in Eq.~(\ref{eq:CIx}). The term of the form (ii) reduces to
\begin{eqnarray}
&&\!\!\!\!\sum_{j k} \Big \langle
|{\cal E}_j|^2
[{\cal D} \ast \Psi_{jA}](t-t_j) ~
[{\cal D} \ast \Psi_{jB}](t+\delta t-t_j) \Big \rangle
\nonumber \\ && \times
\Big \langle |{\cal E}_k|^2
[{\cal D}^\ast\ast\Psi_{kA}^\ast](t-t_k)
~[{\cal D}^\ast\ast\Psi_{kB}^\ast](t+\delta t-t_k)
\Big \rangle.
 \label{eq:type-ii}
\end{eqnarray}
The summations over $j$ and $k$ are seperable and so the term of type (ii) in Eq.~(\ref{eq:CIx}) reduces to a product of 2-point correlation functions, $C^{(2,0)}_{AB}(\delta t) C^{(2,0)\ast}_{AB}(\delta t)$. These functions, defined by Eq.~(\ref{eq:C20}), are equal to zero since they are averages of the correlation of positive-frequency functions. Thus the term of the form (ii) vanishes.

The term of the form (iii) has a similar expression:
\begin{eqnarray}
&&\!\!\!\!\sum_{j k} \Big \langle
|{\cal E}_j|^2
[{\cal D} \ast \Psi_{jA}](t-t_j)
~[{\cal D}^\ast \ast \Psi_{jB}^\ast](t+\delta t-t_j) \Big \rangle
\nonumber \\ && \times
\Big \langle |{\cal E}_k|^2
[{\cal D}^\ast\ast\Psi_{kA}^\ast](t-t_k)
~[{\cal D}\ast\Psi_{kB}](t+\delta t-t_k)
 \Big \rangle.
 \label{eq:type-iii}
\end{eqnarray}
This reduces to $C^E_{AB}(\delta t) C_{AB}^{E\ast}(\delta t)$, which is generally nonzero (on account of the location of the complex conjugates); this is equivalent to Eq.~(8) of \citet{2013arXiv1311.1806L}. Thus Eq.~(\ref{eq:CIx}) reduces to
\begin{equation}
C^{\delta I}_{AB}(\delta t) = |C^E_{AB}(\delta t)|^2 + C^{\delta I}_{AB;j=k=l=m}(\delta t).
\label{eq:CI}
\end{equation}
Here the first term is that of form (iii) and the second is of form (iv). Further computations will focus on this last term that is of interest since we already know that $|C_{AB}(\delta t)|^2$ is independent of dispersion measure. The last term is easily seen to be the connected part of the correlation function, and we will use this nomenclature below.

[Note that in the absence of the connected part $C^{\delta I}_{j=k=l=m}(\delta t)$, the variance of the intensity fluctuation in a channel $C^{\delta I}_{AA}(0)$ is equal to the square of the mean intensity $\bar I_A = C^E_{AA}(0)$. This is appropriate for a complex time series $\bar E_A(t)$, since in the Gaussian limit its ``intensity'' follows a rescaled $\chi^2$ distribution with 2 degrees of freedom. For a real time series with only 1 degree of freedom the disconnected contribution to the variance is twice the mean intensity squared. Mathematically this arises because $\Psi_j$ has only positive frequencies.  A real $\Psi_j$ would result in a factor of $2$ in front of the $|C^E_{AB}(\delta t)|^2$ term because the terms of form (ii) would contribute equally to those of form (iii).]

The connected term is given by
\begin{eqnarray}
C^{\delta I}_{AB; j=k=l=m}(\delta t) 
\!\!\!\! & = &\!\!\!\! \Gamma \int_{-\infty}^\infty d\tau \Big \langle |{\cal E}_j|^4  [{\cal D} \ast \Psi_{jA}](\tau) \times [{\cal D}^\ast \ast \Psi_{jA}^\ast](\tau)
\nonumber \\
& & \times [{\cal D} \ast \Psi_{jB}](\tau+ \delta t) \times [{\cal D}^\ast \ast \Psi_{jB}^\ast](\tau + \delta t) \Big \rangle
\nonumber \\
 & = & \!\!\!\! \Gamma \Big \langle |{\cal E}_j|^4  \big\{ [ {\cal R}| ({\cal D} \ast \Psi_{jA}|^2 ]
 \ast |{\cal D} \ast \Psi_{jB}|^2 \big\} (\delta t)
 \Big\rangle,
 \label{eq:Cjklm}
\end{eqnarray}
where we have used the same trick as in \S\ref{sec:stat-c} to make the integral over $\tau$ in Eq.~(\ref{eq:Cjklm}) a convolution by applying a time reversal operator, ${\cal R}$.  (Our formulae for $C^{\delta I}_{j=k=l=m}$ and $C^E$ double count the case $i=j=k=l$ in a manner that does not matter in the limit that a large number of pulses are contributing at any one time.)

We consider the case of a single filter $A$, and drop its subscript for convenience.  In Fourier space, $C_{j=k=l=m}(\delta t)$ becomes 
\begin{eqnarray}
\tilde C^{\delta I}_{j=k=l=m}(f) \!\!\!\!\!\!\!\!\!\!\!\!\!\!\!\!\!\!\!\!\!\!\!\! &&
\nonumber \\
& = &\!\!\!\! \Gamma \; \bigg \langle  |{\cal E}_j|^4 \int_{-\infty}^\infty \rmd f' \,\tilde{\cal D}(f') \tilde\Psi_j(f') \widetilde{{\cal D}^\ast}(f - f')  \widetilde{\Psi_j^\ast}(f - f')  \nonumber \\
&& \times \int_{-\infty}^\infty \rmd f''\, \widetilde{\cal{RD}}(f'') \widetilde{{\cal R} \Psi_j}(f'') \widetilde{{\cal RD}^\ast}(f - f'')  \widetilde{{\cal R} \Psi_j^\ast}(f - f'')
\bigg \rangle \nonumber \\
& = & \Gamma \; \bigg \langle  |{\cal E}_j|^4 \int_{-\infty}^\infty \rmd f'\, \tilde{\cal D}(f') \tilde\Psi_j(f') \tilde{\cal D}^\ast(-f + f')  \tilde\Psi_j^\ast(f - f') \nonumber \\
&&\times  \int_{-\infty}^\infty \rmd f''\, \tilde{\cal D}(-f'') \widetilde{{\cal R} \Psi_j}(f'') \tilde{\cal D}^\ast(f - f'')  \widetilde{{\cal R} \Psi_j^\ast}(f - f'')
\bigg \rangle \nonumber \\
& = & \Gamma \; \bigg \langle  |{\cal E}_j|^4 \int_{{\mathbb R}^2} \rmd f'\, \rmd f''\, e^{-\rmi [\phi(f') - \phi(-f+f') + \phi(-f'') - \phi(f-f'')]} \nonumber\\
&& \times \tilde\Psi_j(f') \tilde\Psi_j^\ast(f'-f) \tilde \Psi_j(-f'') \tilde \Psi_j^\ast(f - f'')  \bigg \rangle ,
\nonumber \\
& = & \Gamma \; \bigg \langle  |{\cal E}_j|^4 \int_{{\mathbb R}^2} \rmd f'\, \rmd f''\, e^{-\rmi [\phi(f') - \phi(f'-f) + \phi(f'') - \phi(f''-f)]} \nonumber\\
&& \times \tilde\Psi_j(f') \tilde\Psi_j^\ast(f'-f) \tilde \Psi_j(f'') \tilde \Psi_j^\ast(f + f'')  \bigg \rangle ,
\label{eq:Cjklm3}
\end{eqnarray}
where in the second equality we used that $\widetilde{\cal{RD}}(f) =  \tilde{\cal D}(-f)$, $\widetilde{{\cal D}^\ast}(f) =  \tilde{\cal D}^\ast(-f)$, and $\widetilde{\cal {RD}^\ast}(f) =  \tilde{\cal D}^\ast(f)$. The last equality involved a change of variables: $f''\rightarrow-f''$.
%It is not interesting to evaluate this for a $\delta$-function pulse as then the frequency response is infinitely broad.  

Let us evaluate $\tilde C_{j=k=l=m}$ for a broadband source viewed through a Gaussian filter of 1$\sigma$ width $\sigma_f$ and mean $f_A$, with $\sigma_f\ll f_A$. The total normalization is absorbed into ${\cal E}_j$:
\begin{equation}
\tilde\Psi_j(f) = \rme^{-(f-f_A)^2/2\sigma_f^2}.
\end{equation}
We take the phase shift $\phi$ to be quadratic for frequencies near $f_A$, i.e. we use Eq.~(\ref{eq:lin-disp}).
With these assumptions, Eq.~(\ref{eq:Cjklm3}) becomes a 2-dimensional Gaussian integral over frequency, peaked near $(f',f'')\approx (f,f)$. This evaluates to
\begin{equation}
\widetilde C^{\delta I}_{j=k=l=m}(f) = \pi \Gamma \langle|{\cal E}_j|^4\rangle \sigma_f^2  \exp \left[-\frac{f^2(1 + 4\pi^2 D^2 \sigma_f^4)}{2 \sigma_f^2} \right]. \label{eqn:tildeCcon}
\end{equation}
The last factor depends on dispersion.

The  Fourier transform of Eqn~(\ref{eqn:tildeCcon}) is
\begin{equation}
 C^{\delta I}_{j=k=l=m}(\delta t) = \frac{ \pi \Gamma \sigma_f^2 \langle|{\cal E}_j|^4\rangle}{ \sqrt{ (2 \pi \sigma_f^2)^{-1} + 2 \pi D^2 \sigma_f^2 } }
 \exp \left( -\frac{2 \pi^2 \sigma_f^2}{1 + 4 \pi^2 D^2  \sigma_f^4}\delta t^2\right).
\label{eq:cft}
\end{equation}
Compare this to the disconnected part for the assumed waveform
\begin{equation}
|C_{E}(\delta t)|^2 = \pi \Gamma^2 \sigma_f^2  \langle | {\cal E}_j|^4 \rangle \rme^{-2\pi^2 \sigma_f^2\, \delta t^2}.
\label{eq:CEsq}
\end{equation}
Our same reasoning holds as before that the disconnected part is much larger as $\delta t \rightarrow 0$.  It is only at significant temporal lags that the connected part becomes larger than $C_{E}(t)$.

\subsection{Detectability of connected part}
\label{ss:detectability}

Here we estimated the signal-to-noise ($S/N$) at which $C^{\delta I}_{j=k=l=m}(\delta t)$ can be detected
in the most optimistic limit that the synchrotron source dominates the instrumental system temperature.  At large time lags the $S/N$ with which $C_{j=k=l=m}(\delta t)$ can be measured in a given sample (i.e. in a time of order $\sigma_f^{-1}$) is\footnote{This equation assumes that $|D|\sigma_f^2\gg 1$: see Eq.~(\ref{eq:cft}). From Eq.~(\ref{eq:typdisp}), this is trivially satisfied except for extraordinarily narrow filters. If such a narrow filter were used, then we should replace $|D|$ in the estimates below by $\sigma_f^{-1/2}$. Of course this would then give the detectability of the connected correlation function, and not of the dispersion.}
\begin{equation}
\frac{S}{N} =  \frac{|C^{\delta I}_{j=k=l=m}(\delta t)|}{|C_E(0)|^2} \approx  \frac{1}{\sqrt{2 \pi}\, \Gamma |D| \sigma_f}  \rme^{ -\delta t^2/(2 D^2 \sigma_f^2)},
\end{equation}
where the latter approximate equality used Eq.s~(\ref{eq:cft}) and (\ref{eq:CEsq}).
The disconnected part can be measured for $\sim D \sigma_f^2$ temporal lags (the width of the Gaussian $D\sigma_f$, divided by the sample time $\sigma_f^{-1}$) each with $\sigma_f t$ independent samples, meaning the cumulative signal to noise is
\begin{equation}
\left(\frac{S}{N}\right)^2 \sim \frac{\sigma_f t}{\Gamma^2  |D|}.% = 10{-93} \frac{\xi}{1~{\rm GHz}} \frac{t}{1~{\rm yr}} \left(\frac{\Gamma}{10^50 {\rm s^{-1}}}\right)^2  \frac{D}{1~s ~GHz^{-1}}. 
\label{eqn:SNsq}
\end{equation}

To evaluate Eq.~(\ref{eqn:SNsq}), we need an estimate for the number of electrons that contribute at any time.  The number of electrons whose emission is beamed towards an observer from a cosmological synchrotron source can be obtained from the formulae for synchrotron radiation \citep[e.g][\S6.2]{1986rpa..book.....R}: $P_\nu\sim e^3 B/(m_e c^2) F(\nu/\nu_c)$ (units: erg$\,$s$^{-1}\,$Hz$^{-1}$), the function $F(x)$ is peaked near $x\sim 1$ with $F\sim 1$, $\nu_c \sim \gamma^2 e B /(m_e c)$, and that in any instant an electron illuminates $\Omega \sim \gamma^{-2}$. The power radiated per electron per unit frequency is then $P_\nu \sim e^2\nu/(\gamma^2c)$. The number of electrons contributing to the radiation at any one time is then $N_e = \Omega d_{\rm L}^2  f_\nu / P_\nu$, since $\Omega/(4\pi)$ is the fraction of electrons contributing, $4\pi d_{\rm L}^2 f_\nu$ is the total power emitted per unit frequency (in all directions), and $P_\nu$ is the contribution of any one electron. This evaluates to
\begin{equation}
N_e \sim \frac{\Omega d_{\rm L}^2  f_\nu}{P_\nu}
\sim \frac{cd_{\rm L}^2  f_\nu}{e^2\nu}
 \sim 10^{52} \left(\frac{f_\nu}{1~{\rm Jy}} \right) \left( \frac{\nu}{1~\rm GHz} \right)^{-1} \left(\frac{d_L}{1~{\rm Gpc}} \right)^2.
%
% in cgs: c/e^2 = 1/(alpha hbar) = 137/(1e-27 erg*s) = 1e29 /erg/s
% fiducial numbers are 1 Jy = 1e-23 erg/cm2/s/Hz, 1 GHz = 1e9 Hz, 1 Gpc = 3e27 cm
% so 1e29 * 1e-23 * (3e27)^2 / 1e9 = 1e52
%
\end{equation}

Thus, since $\Gamma \sim N_ef$, Eq.~(\ref{eqn:SNsq}) evaluates to
\begin{equation}
\left(\frac SN\right)^2 \sim \frac{t(\sigma_f/f)}{N_e^2 f |D|}
\sim 10^{-97} \frac{t_{\rm yr}(\sigma_f/f)}{(N_e/10^{52})^2f_{\rm GHz} |D_{\rm s/GHz}|}.
\label{eq:stondI}
\end{equation}
The factors of order unity need not be computed here; the signal-to-noise ratio is completely negligible.

This S/N estimate was for the correlation function of $\delta I$.  We could imagine instead measuring $\langle E(t) E^\ast(t+\delta t_1) E(t +\delta t_2) E^\ast (t+ \delta t_3) \rangle$ -- the general four point function of the electric field --, which will increase the number of independent lags from $|D| \sigma_f^2$ to $(|D| \sigma_f^2)^3$.  However, this increase is not comparable to $N_e$ as detection would require.  In addition, we could look at even higher order moments and compare the connected to disconnected part, but it is clear that since $S/N$ is proportional to $\sim N_e^{-n}$ with $n \geq 1/2$, where $1/2$ is for the three point function\footnote{For synchrotron emission, a nonzero odd point function requires an asymmetry in the orientation of the source's  gyrating electrons, such as would occur if the system has nonzero net magnetic flux.}, and other factors simply are insufficient to offset this large number.

\subsection{Comparison to Lieu \& Duan}
\label{ss:lieuduan}

In the \citet{2013ApJ...763L..44L} equation for the waveform (their Eq.~1), DM enters (correctly) as a pure phase, and so the argument that the disconnected part must not depend on dispersion (\S\ref{sec:stat-c}) has to remain valid.  \citet{2013ApJ...763L..44L} calculate the disconnected part of the intensity correlation function at zero lag (which does not depend on DM when temporally averaged; their Eq.~11).  They claim that the {\em timescale} of intensity variations is what depends on DM, and their calculations are based on intensity variations involving sums over 2 electrons -- thus they are indeed calculating the disconnected part. Here we show explicitly that the disconnected part of the variation in the \citet{2013ApJ...763L..44L} calculation is indeed independent of DM if we follow their calculation through to its conclusion in the case of $D\neq 0$. (\citealt{2013ApJ...763L..44L} do not provide the complete calculation.) In particular, we show that the coherence time of intensity fluctuations is $\sim \sigma_t \equiv (2\pi \sigma_f)^{-1}$ regardless of $D$, and is not the width of the measured pulse from a single electron, which is $\sim D\sigma_f$ if $|D|\gg \sigma_t^2$.

In the notation of \citet{2013ApJ...763L..44L}, the observed waveform (equivalent to our Gaussian bandpass, but defining $\Delta\omega \equiv 2\pi\sigma_f$) is\footnote{\citet[Eq. 6]{2013ApJ...763L..44L} are missing a factor of $\rmi$ in front of the phase.}
\begin{equation}
\Psi_j^{\rm LD}(t) = \frac{A}{c} \sqrt{\frac{2 \pi}{1 +\rmi \xi}} \Delta \omega \,
\rme^{-a t^2 + \rmi a\xi t^2 - \rmi \omega_0 t +\rmi \phi_j }.
\label{eq:Psij-LD}
\end{equation}
This is \citet[Eq. 6]{2013ApJ...763L..44L} for the waveform where DM enters via $\xi$, and we have used their parameter definitions
\begin{equation}
a \equiv \frac{ \Delta \omega^2 }{ 2 (1+\xi^2)}
\label{eq:LDa}
\end{equation}
and\footnote{Here we use the notation of \citet{2013ApJ...763L..44L}, where $t_j-t_e$ is the propagation time, and take $L$ to be the distance to the source, so that the propagation time is $L/v_{\rm g}$. We work to lowest order in the intergalactic electron density so that we may approximate $v_{\rm g}\approx c$. The first equality us based on $v_{\rm g}=\rmd\omega/\rmd k$, and the last equality used that $D = (2\pi)^{-1}\rmd(L/v_{\rm g})/\rmd f$.}
\begin{equation}
\xi = \left.\frac{\rmd^2\omega}{\rmd k^2}\right|_{\omega_0}\!\! (\Delta \omega)^2 \frac{t_j-t_e}{c^2}
=
- \left.\frac{\rmd (v_{\rm g}^{-1})}{\rmd \omega}\right|_{\omega_0}\!\! (\Delta \omega)^2 \frac{v_{\rm g}^3L}{c^3}
= - 2\pi D(\Delta \omega)^2.
\end{equation}
The amplitude is contained in $\Psi_j^{\rm LD}$ in the notation of \citet{2013ApJ...763L..44L}, so we may take ${\cal E}_j=1$ for all $j$.
Note also that the time of the event $t_j$ appears in the function $\psi$ in \citet{2013ApJ...763L..44L}, whereas here we include it by using $t-t_j$ as the argument.

We can now recompute the correlation functions using this notation.  The electric field 2-point correlation function $C^E(\delta t)$ is
\begin{equation}
C^E(\delta t) = \Gamma [{\cal R} \Psi^{\rm LD}_j \ast \Psi^{{\rm LD}\ast}_j] (t)  = \frac{2 \pi^{3/2} |A|^2}{c^2}\Gamma \Delta \omega\,
\rme^{-\Delta\omega^2\,\delta t^2/4 - \rmi\omega_0\,\delta t},
\label{eq:CELD}
\end{equation}
and the mean intensity is
\begin{equation}
\bar I = C^E(0) = \frac{2 \pi^{3/2} |A|^2}{c^2}\Gamma \Delta \omega,
\label{eq:LDIbar}
\end{equation}
in agreement with \citet[Eq. 8]{2013ApJ...763L..44L}.

The {\em disconnected} contribution to the intensity fluctuation can be obtained by taking the fluctuating part of the intensity (\citealt{2013ApJ...763L..44L}, Eq. 7),
\begin{equation}
I_1(t) = 2\Re \sum_{j<k} \Psi_j^{\rm LD}(t-t_j) \Psi_k^{{\rm LD}\ast}(t-t_k),
\end{equation}
and finding the correlation function at lag $\delta t$, where $\Re$ denotes the real part. The leading part in the limit where $N_e\gg 1$ is the disconnected part, obtained by summing over all distinct pairs $(j,k)$:
\begin{eqnarray}
C^{\delta I}(\delta t) 
&=& \langle I_1(t) I_1(t + \delta t) \rangle
\nonumber \\
&=& \biggl\langle
 4 \sum_{j<k} \Re [\Psi_j^{\rm LD}(t-t_j) \Psi_k^{{\rm LD}\ast}(t-t_k)]
 \nonumber \\
&& \times\Re [\Psi_j^{\rm LD}(t-t_j+\delta t) \Psi_k^{{\rm LD}\ast}(t-t_k+\delta t)]
 \biggr\rangle.
\end{eqnarray}
Substitution of Eq.~(\ref{eq:Psij-LD}) gives
\begin{eqnarray}
C^{\delta I}(\delta t) \!\!\!\! &=& \!\!\!\! 
\biggl\langle
\frac{16\pi^2 |A|^4}{c^4(1+\xi^2)}
\sum_{j<k} 
\rme^{-a^2[(t-t_j/2-t_k/2+\delta t/2)^2 + \delta t^2 + (t_k-t_j)^2]}
\nonumber \\ && \!\!\!\! \times
\cos \left\{ -a\xi[ (t-t_j)^2 - (t-t_k)^2]
+ \omega_0 t_{jk} + \phi_{jk} \right\}
\nonumber \\ && \!\!\!\! \times
\cos \left\{ -a\xi[ (t-t_j+\delta t)^2 - (t-t_k+\delta t)^2]
+ \omega_0t_{jk} + \phi_{jk} \right\}
\biggr\rangle,
\nonumber \\&&
\label{eq:LD-temp.1}
\end{eqnarray}
where $\phi_{jk} \equiv \phi_j-\phi_k$ and $t_{jk}=t_j-t_k$, and we have completed the square in the Gaussian envelope.

It is readily seen that the Gaussian envelope in Eq.~(\ref{eq:LD-temp.1}) can have a nonzero contribution only when $|\delta t|\lesssim a^{-1/2}$, and hence it is tempting to conclude that the coherence time is now $\sim a^{-1/2}$, which grows with $|D|$. But this is not so: a more accurate conclusion from Eq.~(\ref{eq:LD-temp.1}) is that the coherence time must be $\lesssim a^{-1/2}$. To see how this works, one must actually perform the average in Eq.~(\ref{eq:LD-temp.1}). Let us perform the average over phases $\phi_{jk}$ first, using the rule that\footnote{This follows trivially from the product-to-sum rule.}
\begin{equation}
\langle \cos(\alpha+\phi_{jk}) \cos(\beta+\phi_{jk}) \rangle = \frac12\cos(\alpha-\beta).
\end{equation}
Then with some algebraic simplification, Eq.~(\ref{eq:LD-temp.1}) reduces to
\begin{eqnarray}
C^{\delta I}(\delta t) \!\!\!\! &=& \!\!\!\! 
\frac{8\pi^2 |A|^4}{c^4(1+\xi^2)}
\biggl\langle
\sum_{j<k} 
\rme^{-a^2[(t-t_j/2-t_k/2+\delta t/2)^2 + \delta t^2 + t_{jk}^2]}
\nonumber \\ && \!\!\!\! \times
\cos ( 2a\xi t_{jk} \delta t )
\biggr\rangle.
\label{eq:LD-temp.2}
\end{eqnarray}
The key to the coherence time is the cosine factor. At zero lag ($\delta t=0$), the cosines in Eq.~(\ref{eq:LD-temp.1}) are identical, they can be replaced with a $\cos^2$ as in \citet[Eq. 10]{2013ApJ...763L..44L}, and their average value is simply $\frac12$. However, when we look at nonzero coherence times, there is an oscillatory factor: if $t_{jk}$ is an integer multiple of $\pi/(a\xi\,\delta t)$, then the argument of the cosine is an integer multiple of $2\pi$ and the two electrons $j$ and $k$ cause positively correlated intensity fluctuations at $t$ and at $t+\delta t$. However, if $t_{jk}$ is a half-integer multiple of $\pi/(a\xi\,\delta t)$, then that pair of electrons causes negatively correlated intensity fluctuations. This oscillatory or ``fringing'' behaviour has a simple interpretation for highly dispersed pulses ($\xi\gg 1$): if one observes electrons $j$ and $k$ simultaneously, then due to dispersion, the pulses have propagated through a time that differs by $t_{jk}$ and hence their frequencies differ by $Dt_{jk}$. If $Dt_{jk}\delta t$ is an integer then the relative phase of the pulses from $j$ and $k$ is identical at $t$ and $t+\delta t$, and so that pair of electrons either interferes constructively at both $t$ and $t+\delta t$ or destructively at both $t$ and $t+\delta t$. But if $Dt_{jk}\delta t$ is a half-integer then the relative phase changes by $\pi$, and so that pair of electrons interferes constructively at either $t$ or $t+\delta t$ and destructively at the other. The {\em intensity} correlation function at lag $\delta t$ thus receives a positive or negative contribution.

This means that the intensity fluctuations can become uncorrelated either if $|\delta t|\gg a^{-1/2}$ {\em or} if the range $\sim a^{-1/2}$ of $t_{jk}$ allowed by the Gaussian envelope contains many fringe periods, i.e. is $\gg 1/(a\xi\,\delta t)$. The latter condition is
\begin{equation}
a^{-1/2} \gg \frac{1}{a\xi\,\delta t} ~~~\rightarrow~~~
\delta t \gg \frac{1}{a^{1/2}\xi} \sim \frac{\sqrt{1+\xi^{-2}}}{\Delta \omega}.
\end{equation}
Thus we expect the coherence time to be the minimum of $a^{-1/2} \sim \sqrt{1+\xi^2}\,/\Delta\omega$ and $\sqrt{1+\xi^{-2}}\,/\Delta\omega$, i.e. it should be of order $\Delta\omega^{-1}$, regardless of the dispersion parameter $\xi$.

To complete our computation of the intensity fluctuation correlation function, we replace the summation in Eq.~(\ref{eq:LD-temp.2}) with an integral over $t_j$ and $t_k$ (and include a prefactor of the rate $\Gamma^2$). In fact it is easiest to switch variables to $\tau \equiv t-(t_j+t_k)/2+\delta t/2$ and $t_{jk}$; the Jacobian is unity. Moreover, the region of integration is over ${\mathbb R}^2$, with a factor of $\frac12$ to avoid double-counting pairs. The result is
\begin{eqnarray}
C^{\delta I}(\delta t) \!\!\!\! &=& \!\!\!\! 
\frac{4\pi^2 |A|^4}{c^4(1+\xi^2)} \Gamma^2 \rme^{-a^2\,\delta t^2}
\int_{-\infty}^\infty \rmd\tau\,\rme^{-a^2\tau^2}
\nonumber \\ && \times
\int_{-\infty}^\infty\rmd t_{jk}\,
\rme^{-a^2 t_{jk}^2}
\cos ( 2a\xi t_{jk}\, \delta t)
.
\label{eq:LD-temp.3}
\end{eqnarray}
The Gaussian ($\tau$) and Gaussian-oscillatory ($t_{jk}$) integrals are easily evaluated as $\sqrt\pi\,/a$ and $(\sqrt\pi\,/a)\rme^{-a\xi^2\delta t^2}$ respectively. The result is
\begin{equation}
C^{\delta I}(\delta t) = \frac{4\pi^3 |A|^4}{c^4a(1+\xi^2)} \Gamma^2 \rme^{-a^2\,\delta t^2} \rme^{-a\xi^2\delta t^2}
= \bar I^2 \rme^{-\Delta\omega^2\,\delta t^2/2},
\end{equation}
where in the second equality we have used Eq.~(\ref{eq:LDa}) for $a$ and Eq.~(\ref{eq:LDIbar}) for $\bar I$, and finally note that since we were calculating the disconnected part $C^{\delta I}(\delta t) = |C^{E}|(\delta t)^2$, where $C_E$ is given by Eq.~(\ref{eq:CELD}).

%Then the intensity correlation function is given by Eq.~(\ref{eq:CI}):
%\begin{equation}
%C^{\delta I}(\delta t;{\rm disconnected}) = \frac{|C^E(\delta t)|^2}{\bar I^2} = \rme^{-\Delta\omega^2\,\delta t^2/2}.
%\label{eq:IC:LD}
%\end{equation}
%This is indeed independent of the dispersion measure.

\subsection{Quantum mechanics and intensity fluctuations}
\label{ss:QM}

The discussion thus far has been entirely classical, and it has been shown that the connected part of the intensity correlation function, which is suppressed by $1/N_e$ relative the to the disconnected part, is the only part that is sensitive to DM. On the other hand, \citet{2013arXiv1311.1806L} have argued that only the electrons whose emitted photons reach the observer should be counted in this calculation. In their picture, in the limit where other noise sources are negligible (i.e. where the source is much brighter than the sky integrated over the beam width and with no thermal emission in the instrument), the $S/N$ at which the connected part of the intensity correlation function can be detected is $\propto 1/\bar n_\gamma$ in place of $1/N_e$ in our expressions, where $\bar n_\gamma$ is the source photon occupation number (see discussion in their \S4).  Since bright radio sources contribute brightness temperatures of $T_b \sim 1$K in the beam of a radio instrument, they have $\bar n_\gamma \sim 20 \nu_{\rm GHZ}^{-1} T_{b \rm K}$, which is much smaller than $N_e \sim 10^{50}$.   This motivates the question of how our classical derivations are modified by quantum mechanics. The basic question is as follows: are the quantum (Poisson) intensity fluctuations a feature of the source that can be dispersed by passage through a plasma like any classical intensity fluctuation, as \citet{2013arXiv1311.1806L} assumed?\footnote{The calculation in \citet{2013arXiv1311.1806L} is actually classical; while incorporating the Poisson noise term by constructing their signal as a sum of pulses, their signal ``$\Phi(t)$'' is a number, not an operator.} Or is the Poisson noise somehow immune to dispersion due to its quantum nature?

It is shown in Appendix~\ref{app:QM} that -- within the full context of quantum field theory, and with appropriate approximations -- the density matrix of the received radiation is equivalent to a statistical superposition of coherent states. These states are displaced vacuum states, i.e. states in which the electromagnetic field operators $\hat{\bmath A}({\bmath x})$, $\hat{\bmath E}({\bmath x})$, etc. are equal to a classical field solution, plus the operators corresponding to the quantum vacuum fields. Their properties are described in detail in \S\ref{ss:setup-i}. The statistical distribution that goes into these coherent states is merely the classical waveform emitted by the electrons, averaged over a randomly chosen phase in their orbits. It is thus apparent that the full quantum state of the received radiation can contain no information that would not be present classically. In particular, the intensity fluctuations measured by a photoelectric detector, which correspond to a normal-ordered 4-point correlation function of the electric field \citep[Ch. 12]{1995ocqo.book.....M}, are equal to the classical correlation function computed from the coherent state amplitudes according to the optical equivalence theorem \citep{1963PhRvL..10..277S} -- i.e. those computed in the preceding sections.  This result formally settles the question in favour of the classical analysis presented herein: the intensity correlation function is independent of DM up to corrections of order $1/N_e$.

To understand intuitively why this is so, one must consider the nature of Poisson noise for identical bosonic particles. Consider a single-mode coherent state of complex amplitude $v$. A coherent state, or displaced vacuum state, is a right eigenstate of the annihilation operator $\hat a$ with eigenvalue $v$ (and {\em not} a right eigenstate of $\hat a^\dagger$ with eigenvalue $v^\ast$). It is for this reason that the optical equivalence theorem for coherent states applies to normal-ordered correlation functions: e.g. $\bar n_\gamma \equiv \langle \hat N\rangle = \langle \hat a^\dagger \hat a\rangle = |v|^2$ and $\langle \hat a^{\dagger2} \hat a^2 \rangle = |v|^4$. The variance of the photon occupation number is given by
\begin{equation}
\langle \hat N^2 \rangle - \langle \hat N\rangle^2 = \langle \hat a^\dagger \hat a \hat a^\dagger \hat a \rangle - |v|^4 = 
\langle \hat a^\dagger [\hat a, \hat a^\dagger ]\hat a \rangle
= \bar n_\gamma;
\end{equation}
this is in fact the Poisson noise term, and it arose entirely from operator-ordering considerations (the square of the photon number operator $\hat N=\hat a^\dagger \hat a$ is not normal-ordered). Since the phase delay due to dispersion applies to the coherent state amplitudes, and not directly to numbers of photons, this added Poisson noise is not dispersed.

To frame this discussion in the context of a full statistical distribution, the number of photons in that mode is Poisson-distributed with mean $|v|^2$; in a statistical mixture of such states where $v$ is complex Gaussian distributed with zero mean and variance $\bar n_\gamma$, then the occupation number is Bose-Einstein distributed, with mean $\bar n_\gamma$ and variance $\bar n_\gamma^2+\bar n_\gamma$ (here $\bar n_\gamma^2$ is the classical variance, and $\bar n_\gamma$ is the additional contribution from Poisson noise). This is a good model for the photon density matrix in a single mode of the electromagnetic field in the limit of a large number of emitting electrons $N_e$, since then the central limit theorem will force the amplitudes $v$ to have a Gaussian distribution. It is to this underlying semiclassical Gaussian field that the machinery in the previous sections should be applied.

% I took out the 3 point function -- C.H.

\section{Masers as sources}
\label{sec:maser}

The previous sections showed that dispersion is undetectable for incoherent sources of radiation because the radiation is highly Gaussian.  A class of extragalactic sources exist where the emission is stimulated and potentially less Gaussian:  the 1.6\;GHz OH and 22\;GHz H$_2$O ``mega-masers'' \citep[e.g.][]{2005ARA&A..43..625L}.  %Mega-masers have been detected out to redshifts of a few tenths.  % typically found in luminous infrared galaxies and the nuclear regions of AGN.
  Astrophysical masers do not have a well-defined cavity like a laboratory laser.  Instead, maser radiation is broadband in nature, without the phase coherence of a laboratory laser.  %(with the line width set by the temperature, velocity gradients, and amount of amplification) and 
    
Still, maser radiation may exhibit correlations between non-equal frequencies and hence non-Gaussianity \citep{1978PhRvA..17..701M, 1984MNRAS.211..799F, 2009MNRAS.399.1495D}.  Saturation of the population inversion and as a result in the growth of the electric field occurs first in modes that have higher than average amplitudes.  It acts to reduce the variance of the intensity, $C^{\delta I}(\delta t = 0)$, relative to the Gaussian expectation.  Simple one-dimensional calculations with parameters motivated by Galactic masers find a connected contribution to $\tilde C^{\delta I}(f)$ that is $\sim-10\%$ the disconnected part at line center \citep{2009MNRAS.399.1495D}, although there are many effects that may reduce the connected term relative to these predictions.  This level of non-Gaussianity is comparable to the observational limit on the connected part of $C^{\delta I}(0)$ in Galactic OH masers \citep{1972PhRvA...6.1643E}.  %The non-Gaussian contribution to $C^{\delta I}(0)$ has been constrained to be less than $1\%$ in Galactic OH masers \citep{1972PhRvA...6.1643E} and $5\%$ on H$_2$O masers \citep{1981BAAS...13..508M}.

However, the connected contribution to $C^{\delta I}(\delta t )$ in a maser owes its existence to correlations between modes separated by on the order of the homogeneous line width of the masing molecules, $\Delta f_{\rm hom}$ \citep{2009MNRAS.399.1495D}.  In order for plasma dispersion to have a significant impact on the received $C^{\delta I}(\delta t)$, we require $|D| \Delta f_{\rm hom}^2 \gtrsim 1$.   Taking parameters appropriate for OH masers at $z\sim 1$, we find $|D| \Delta f_{\rm hom}^2 \sim 10^{-10}$ assuming DM$=10^3\,$pc$\,$cm$^{-3}$, and even smaller values result for H$_2$O masers.  Therefore, the connected part of maser emission is not significantly altered by dispersion.
% this criterion is far from the being satisfied for both OH ($\Delta f_{\rm hom} \sim 0.1$ Hz) and H$_2$0 ($\Delta f_{\rm hom} \sim ???$ Hz) masers an

\section{Discussion}
\label{sec:disc}

The measurement of a dispersion measure to a cosmological radio source would open up a new window on the study of intergalactic baryons. Since all confirmed cosmological radio sources are constant over timescales much larger than those affected by dispersion, the temporal delay in the arrival of different frequencies cannot be used to measure dispersion unlike for Galactic pulsars.  However, recently several schemes have been proposed to measure the dispersion to such continuous sources using (1) the 2-point correlation function of the electric field, i.e. the delay of a lower-frequency channel relative to a higher-frequency channel \citep{2013MNRAS.433.2275L}; (2) the 2-point correlation function of intensity fluctuations, which should have a longer timescale due to the spreading of arrival time within a given frequency channel \citep{2013ApJ...763L..44L}; and (3) the quantum corrections to the intensity fluctuations, i.e. dispersion of the Poisson noise fluctuations in the source intensity \citep{2013arXiv1311.1806L}. Our analysis has shown that, due to various subtleties, none of these methods work. Indeed, under very general assumptions, the observed signal from a continuous point source has no information beyond the 2-point correlation function of the electric field (which evaluated at zero lag yields the intensity and polarization Stokes parameters).  This result is a consequence of the central limit theorem and the large number of incoherently emitting electrons contributing to the observed waveform for any astrophysical source.

The possibility of measuring cosmological dispersion measures remains enticing, but will only be possible using sources that are time-variable or coherent (and hence potentially non-Gaussian). Masers are coherent, but for typical parameters the dispersion across their very narrow line widths is too small. Thus, given the calculations presented herein, the availability of DM measurements as a probe of the intergalactic medium remains contingent on the cosmological interpretation of the fast radio bursts \citep{2013Sci...341...53T} or the existence of some similar class of fast radio transients.

\section*{Acknowledgments}

We thank Eric Huff, Richard Lovelace, and David Weinberg for useful comments.

During the preparation of this paper, CH has been supported by the US Department of Energy under contract DE-FG03- 02-ER40701, the David and Lucile Packard Foundation, the Simons Foundation, and the Alfred P. Sloan Foundation.
MM acknowledges support by the National Aeronautics and Space Administration
through Hubble Postdoctoral Fellowship awarded by the Space
Telescope Science Institute, which is operated by the
Association of Universities for Research in Astronomy,
Inc., for NASA, under contract NAS 5-26555.

\appendix

\section{Quantum mechanical treatment of synchrotron pulses}
\label{app:QM}

In the main text, we have treated the emission of synchrotron radiation classically. In particular, \citet{2013arXiv1311.1806L} argued that the rate of pulses $\Gamma$ appearing in the correlation function formulae should be the number of {\em photons received} per unit time $\dot N_{\rm ph-rec}$, whereas in our classical treatment it is the number of {\em electrons whose synchrotron beams sweep over the observer} per unit time $\dot N_{\rm el-beam}$.\footnote{\citet{2013arXiv1311.1806L} use the symbol $\lambda$ for this rate instead of $\Gamma$ here and in \citet{2013MNRAS.433.2275L}.} In practical situations, $\dot N_{\rm ph-rec}\ll \dot N_{\rm el-beam}$. Since the dimensionless connected part of the intensity fluctuation $C_{j=k=l=m}(\delta t)/\bar I^2$ is proportional to $1/\Gamma$, and it is this part that is sensitive to DM, it is important that we resolve this issue. Is the classical calculation correct, so that $\Gamma$ is large and the connected intensity fluctuations are tiny, as found in the main text? Or should only the photons that are received by the observer contribute to $\Gamma$, as claimed by \citet{2013arXiv1311.1806L}, resulting in much larger connected intensity fluctuations?

The purpose of this appendix is to resolve this issue with a quantum mechanical calculation of an appropriately simplified problem. \citet{1963PhRv..131.2766G} showed that a {\em classical} current source coupled to a quantized radiation field initially in the vacuum state produces a ``coherent'' photon state, i.e. one obtained from the vacuum state by displacing the wave function with the displacement given by the classical field amplitude. We consider here a fully quantized source, and show that with suitable approximations the outgoing photon state is a statistical superposition of coherent states, each corresponding to the classical field amplitude from electrons with a random distribution of phases in their orbits. The conclusion is that the received radiation field is in fact statistically indistinguishable from the classical field with appropriate measurement noise (including the familiar $\frac12hf$ per mode in the case of a coherent receiver) added in. Therefore, quantum corrections to the received electric field do not provide additional information that would be inaccessible classically; in particular they do not add any information about the dispersion measure.

Throughout we use the c.g.s. unit system and the Schr\"odinger picture of quantum mechanics.

\subsection{Assumptions}

Consider the problem of a synchrotron-emitting cloud. For simplicity, we will take the cloud to be optically thin. We will furthermore ignore processes that create or destroy electrons (we are interested only in the synchrotron radiation), ignore the electron spin, and will assume the cloud to be sufficiently dilute that the identical nature of the electrons can be neglected (i.e. we can number them $1...{\cal N}$, and treat their motion as independent degrees of freedom, ignoring wave function anti-symmetrization or state blocking). Of these assumptions, only the optically thin condition is likely to be violated in a realistic AGN.

We further assume here that the magnetic field configuration permits separation of variables. This is not likely to be true in an actual source, but since our only use of this assumption is to construct wave packets in action-angle space (instead of in position-momentum space, which would lead to a much more extended formalism) we do not think it is of critical importance. In particular, we follow each wave packet through only a portion of an orbit, so we expect that exact integrability (or not) would not affect the conclusions. Also the semi-classical limit will be taken in which all quantum numbers are large, here meaning that the change in quantum number in emission of a single photon is small compared with the quantum number itself.

\subsection{Formalism and Hilbert space}

The electromagnetic field is quantized as a wave, with a set of discrete indices $\alpha$ for the various modes and a continuous index $k\in{\mathbb R}^+=(0,\infty)$ describing the wave number.\footnote{The formulae given here for the quantized electromagnetic field are standard; we have used those of \citet[\S10.3]{1995ocqo.book.....M}, and converted them to a continuous $k$-index.} There are many possible choices of mode with one continuous index (e.g. spherical waves, where $k$ is a continuous and the discrete quantum numbers are angular momentum $jm$ and electric or magnetic type parity E or M) but we do not specify these yet. The appropriately normalized transverse radiative magnetic vector potential operator is then given by
\begin{equation}
{\bmath A}^{\rm(rad)}({\bmath x}) = \sum_{\alpha} \int_0^\infty \frac{\rmd k}{2\pi}\, 
\sqrt{\frac{4\pi \hbar c}{k}}
\,\left[ {\bmath Z}_\alpha({\bmath x};k) \hat a_\alpha(k) + 
{\bmath Z}^\ast_\alpha({\bmath x};k) \hat a^\dagger_\alpha(k) \right],
\label{eq:Arad}
\end{equation}
where $\hat a_\alpha(k)$ is an annihilation operator, $\hat a^\dagger_\alpha(k)$ is a creation operator, and the mode functions ${\bmath Z}_\alpha({\bmath x};k)$ are complete over the space of divergenceless vector fields. They obey the orthonormality relation\footnote{Only a 1-dimensional $\delta$-function appears here since $k$ is taken to be a number rather than a vector; the directional dependence is captured in the discrete indices, which may be e.g. angular momentum indices.}
\begin{equation}
\int {\bmath Z}^\ast_\alpha({\bmath x};k) \cdot {\bmath Z}_\beta({\bmath x};k') \,\rmd^3{\bmath x} = 2\pi \delta_{\alpha\beta} \delta(k-k')
\end{equation}
and the eigenvalue equation $\nabla^2{\bmath Z}_\alpha({\bmath x};k) = -k^2{\bmath Z}_\alpha({\bmath x};k)$.
The annihilation operators mutually commute, but have a nontrivial commutation with the creation operators
\begin{equation}
[ \hat a_\alpha(k), \hat a^\dagger_\beta(k')] = 2\pi \delta_{\alpha\beta} \delta(k-k').
\end{equation}
The Hamiltonian is -- aside from an irrelevant additive constant --
\begin{equation}
\hat H_{\rm rad} = \sum_\alpha  \int_0^\infty \frac{\rmd k}{2\pi}\, \hbar ck \hat a^\dagger_\alpha(k) \hat a_\alpha(k).
\end{equation}
The radiative part of the electric field is conjugate to the magnetic vector potential:
\begin{equation}
{\bmath E}^{\rm(rad)}({\bmath x}) = \sum_{\alpha} \int_0^\infty \frac{\rmd k}{2\pi}\, 
\frac{\sqrt{4\pi c \hbar k}}{\rmi}
\, \left[ {\bmath Z}_\alpha({\bmath x};k) \hat a_\alpha(k) -
{\bmath Z}^\ast_\alpha({\bmath x};k) \hat a^\dagger_\alpha(k) \right].
\end{equation}

Our next interest is in the electrons. Since we are neglecting spin, we take these to be described by a complex scalar wave equation in a time-independent background magnetic vector potential ${\bmath A}^{\rm(bg)}({\bmath x})$. The Lagrangian is
\begin{equation}
\hat L_{\rm el} 
= \int_{{\mathbb R}^3} \rmd^3{\bmath x}\,\frac12\Bigl\{ \hbar^2|\dot{\hat\psi}({\bmath x})|^2
- | [-\rmi c\hbar\nabla - e{\bmath A}^{\rm(bg)}]\psi({\bmath x}) |^2
 - m_{\rm e}^2c^4|\hat\psi({\bmath x})|^2
\Bigr\}.
\end{equation}
As usual without an electric potential, the conjugate momentum is $\hat\pi({\bmath x}) = \hbar^2\hat{\dot\psi}^\ast({\bmath x})$ and the Hamiltonian $\hat H_{\rm el}$ is equal to $\hat L_{\rm el}$ but with a $+$ instead of a $-$ in the second two terms.
The complex wave equation is separable as $\psi({\bmath x}) = \phi_{\bmath n}({\bmath x}) \rme^{-\rmi\omega_{\bmath n} t}$, 
where the mode functions are given by
\begin{equation}
\hbar^2\omega_{\bmath n}^2 \phi_{\bmath n}({\bmath x}) = m_{\rm e}^2c^4\phi_{\bmath n}({\bmath x}) +
[-\rmi c\hbar\nabla - e{\bmath A}^{\rm(bg)}({\bmath x})]^2 \phi_{\bmath n}({\bmath x}).
\label{eq:mode-function}
\end{equation}
As this is an eigenvalue equation with eigenvalue $\omega_{\bmath n}^2$ and a positive-definite Hermitian right-hand side, we may choose the $\phi_{\bmath n}({\bmath x})$ to be $L^2$-orthonormal: $\int_{{\mathbb R}^3} \phi_{\bmath n}({\bmath x}) \phi^\ast_{{\bmath n}'}({\bmath x}) \,\rmd^3{\bmath x} = \delta_{{\bmath {nn}}'}$, and complete: $\sum_{\bmath n} \phi_{\bmath n}({\bmath x}) \phi^\ast_{\bmath n}({\bmath y}) = \delta^{(3)}({\bmath x}-{\bmath y})$. The ``electron'' wave operator and its conjugate momentum may then be written as
\begin{equation}
\hat\psi({\bmath x}) = \sum_{\bmath n} 
\frac1{\sqrt{2\hbar\omega_{\bmath n}}} \,
(\hat b_{\bmath n} + \hat d_{\bmath n}^\dagger)
\phi_{\bmath n}({\bmath x})
\end{equation}
and
\begin{equation}
\hat\pi({\bmath x}) = \sum_{\bmath n} 
-\rmi\sqrt{\frac{\hbar^3\omega_{\bmath n}}{2}}\,
(\hat b_{\bmath n} - \hat d_{\bmath n}^\dagger)
\phi_{\bmath n}({\bmath x}),
\end{equation}
where the sum is over $\omega_{\bmath n}>0$. It is readily verified that these operators obey the proper commutation relations with $\hat b_{\bmath n}$ and $\hat d_{\bmath n}$ interpreted as annihilation operators for independent quantum harmonic oscillators (for the particle and antiparticle), and with a Hamiltonian
\begin{equation}
\hat H_{\rm el} = \sum_{\bmath n} \hbar\omega_{\bmath n} (\hat b_{\bmath n}^\dagger\hat b_{\bmath n} + \hat d_{\bmath n}^\dagger \hat d_{\bmath n}),
\end{equation}
again with an irrelevant constant subtracted off. While the antiparticle operators are necessary for the overall consistency of the theory, none of our operations will involve states with antiparticles and so in what follows we suppress terms involving $\hat d_{\bmath n}$.

The interaction of matter and radiation to first order in the radiation amplitude is described by
\begin{equation}
\hat H_{\rm int}^{\rm 1st} = e \int \hat{\bmath A}^{\rm(rad)} \cdot 
\left\{
-\rmi\hbar c\hat\psi^\dagger\nabla\hat\psi + \rmi\hbar\hat\psi\nabla\hat\psi^\dagger + 2e{\bmath A}^{\rm(bg)}\hat\psi^\dagger\hat\psi
\right\} \,\rmd^3{\bmath x};
\end{equation}
the term responsible for emission of synchrotron radiation then has the form
\begin{eqnarray}
\hat H^{\rm(I)}_{\rm int} 
\!\!\!\! &=& \!\!\!\! e \sum_{{\bmath {nn}}'\alpha}\int \frac{\rmd k}{2\pi}\,
\sqrt{\frac{\pi c}{\hbar k\omega_{\bmath n}\omega_{{\bmath n}'}}} \hat a^\dagger_\alpha(k) \hat b_{\bmath n} \hat b^\dagger_{{\bmath n}'}
\int \rmd^3{\bmath x}
\,
{\bmath Z}^\ast_\alpha({\bmath x};k)
\nonumber \\ && 
\cdot \Bigl\{
\rmi \hbar c\phi_{{\bmath n}'}^\ast\nabla\phi_{{\bmath n}}({\bmath x})
- \rmi\hbar c\phi_{\bmath n}\nabla\phi^\ast_{{\bmath n}'}({\bmath x}) + 2e{\bmath A}^{\rm(bg)}
\phi^\ast_{{\bmath n}'}\phi_{\bmath n}({\bmath x})
\Bigr\}.
\label{eq:HI}
\end{eqnarray}
The Hermitian conjugate $\hat H^{{\rm(I)}\dagger}_{\rm int}$ is also present. A term $\hat H^{\rm(II)}_{\rm int}$ containing two factors of the radiation Hamiltonian is also present, but we do not need its explicit form.
The total interaction Hamiltonian is thus $\hat H_{\rm int} = \hat H^{\rm(I)}_{\rm int} + \hat H^{{\rm(I)}\dagger}_{\rm int} + \hat H^{\rm(II)}_{\rm int}$.

The relevant Hilbert space thus consists of the photon and electron degrees of freedom.

\subsection{Setup of the problem; initial conditions}
\label{ss:setup-i}

In the synchrotron emission problem, the initial state of the electromagnetic field is usually taken to be the vacuum, $|{\rm vac}\rangle$. We will be slightly more general here in order to derive results that are useful later, and choose the {\em coherent state} \citep{1963PhRv..130.2529G,1963PhRv..131.2766G} $|v_\alpha(k)\rangle$, defined by
\begin{equation}
|v_\alpha(k)\rangle = \exp \Bigl\{ \sum_\alpha \int \frac{\rmd k}{2\pi} [ v_\alpha(k) \hat a^\dagger_\alpha(k) - v_\alpha^\ast(k) \hat a_\alpha(k) ] \Bigr\}
|{\rm vac}\rangle.
\label{eq:def-coherent}
\end{equation}
It is important to note that a coherent state for a photon {\em field} is labeled by a set of complex functions $v_\alpha(k)$ for each mode. The operator in brackets is an anti-Hermitian linear combination of the generalized coordinate operators $\hat{\bmath A}$ and generalized momentum operators $\hat{\bmath E}$. Thus the coherent state can be thought of as a displaced vacuum state: if $v_\alpha(k)$ is real, then the complex exponential in Eq.~(\ref{eq:def-coherent}) is a displacement operator in the coordinate-space representation of the wave function; if $v_\alpha(k)$ is purely imaginary, then it is a displacement operator in the momentum-space representation of the wave function.

A general discussion of coherent states and their properties can be found in \citet[\S11]{1995ocqo.book.....M}. The most important properties are as follows:
\begin{list}{$\bullet$}{}
\item An annihilation operator acting on a coherent state returns the state's value,
$\hat a_\alpha(k)|v_\beta(k')\rangle = v_\alpha(k) |v_\beta(k')\rangle$.
\item Any density matrix on the photon space can be represented formally as a statistical superposition of coherent states (though not necessarily with positive weight).
\item In a coherent state, the expectation value of any normal-ordered operator $\hat a^\dagger_{\alpha_1}(k_1) ... \hat a_{\alpha_M}(k_M)$ is obtained by replacing $\hat a^\dagger_\alpha(k) \rightarrow v^\ast_\alpha(k)$ and $\hat a_\alpha(k) \rightarrow v_\alpha(k)$ (the optical equivalence theorem).
\item Finally, the coherent state is not an eigenstate of the free Hamiltonian, but it does evolve simply as
$\exp(-\rmi \hat H_{\rm rad}t/\hbar) |v_\alpha(k) \rangle = | \rme^{-\rmi ckt} v_\alpha(k) \rangle$,
where the (irrelevant) zero-point energy of the radiation Hamiltonian has been removed.
\end{list}

\subsection{WKB approximation and correspondence principle}

The absorption and emission of radiation by charged particles in a potential with a separable Hamiltonian has a long history but is not often covered in standard texts. Pioneering work, predating quantum field theory, can be found in \citet{1924PhRv...24..330V, 1924PhRv...24..347V}. A full derivation of the results is given here however, to be consistent with the formalism of quantum field theory and the notation used elsewhere in this appendix.
We use the Wentzel-Kramers-Brillouin (WKB) or eikonal approximation to the solutions to the wave equation with classical Hamiltonian
\begin{equation}
H_{\rm cl}({\bmath x},{\bmath p}) = \sqrt{m_{\rm e}^2c^4 + c^2[({\bmath p}-e{\bmath A}^{\rm(bg)}(x)]^2}.
\end{equation}

The WKB approximation to the stationary states can be formulated in terms of action-angle variables. In cases where the quantum-mechanical wave equation is separable, the classical Hamilton-Jacobi equation is also separable and hence one can construct a set of conjugate action-angle variables, the angles $\{\theta_\mu\}_{\mu=1}^3$ (periodic over the domain from 0 to $2\pi$, and here taken to be a function of the actions and the spatial coordinates) and the actions $\{\hbar {n}_\mu\}_{\mu=1}^3$. The solution $W({\bmath x},{\bmath n})$ to the Hamilton-Jacobi equation, i.e. $\hbar\omega_{\bmath n} = H_{\rm cl}(x_a, \partial W/\partial x_a)$ is Hamilton's characteristic function and has units of action. It is the generating function for the canonical transformation from $(x_a,p_a) \rightarrow (\hbar n_\mu, \theta_\mu)$. The relation is given explicitly by
\begin{equation}
p_a =\left. \frac{\partial W}{\partial x_a}\right|_{{\bmath x},{\bmath n}}
~~~{\rm and}~~~
\theta_\mu = \hbar^{-1} \left.\frac{\partial W}{\partial n_\mu}\right|_{{\bmath x},{\bmath n}};
\label{eq:direct1}
\end{equation}
see e.g. \citet[\S9--3]{G80}. The mode frequencies are related to the classical Hamiltonian by $\hbar\omega_{\bmath n} = H_{\rm cl}$. The Hamiltonian is a smooth function of the quantum numbers for a separable system; the three classical fundamental frequencies are given by $\hbar\Omega_\mu=\partial H_{\rm cl}/\partial n_\mu$. Since there are three actions (and hence three quantum numbers) it is convenient to index the states by the triplet of integers ${\bmath n}\in{\mathbb Z}^3$; thus the index for the electron mode functions $\phi_{\bmath n}$ will henceforth be written in boldface.

In a one-dimensional quantum mechanical system (where $x$, $p$, $\theta$, and $n$ have only one component and so are written as scalars), the WKB solution may be written as
\begin{equation}
\phi_{n}^{\rm 1~d.o.f.}(x) = {\mathfrak C} \sum_{\rm streams} \frac1{|\partial x/\partial \theta|^{1/2}} \rme^{\rmi W(x,n)/\hbar},
\label{eq:1dof}
\end{equation}
where the summation is over the different streams, i.e. the different possible values of momentum (or angle) at fixed action $\hbar n$ and position $x$. (In textbook examples, there are usually two streams, one with positive velocity and one with negative velocity.) The prefactor ${\mathfrak C}$ does not depend on position. The characteristic function $W(x,n)$ is normally written in quantum mechanics texts as the ``action'' $\int p\,\rmd x$, but by Eq.~(\ref{eq:direct1}) the momentum is equal to $\partial W/\partial x$ and so these forms are equivalent. The denominator is normally written as the classical velocity $\dot x_{\rm cl}$ in quantum mechanics texts, but it may also be written as $\partial x/\partial\theta$ since the conversion factor $\dot\theta_{\rm cl} = \Omega$ is independent of $x$. The requirement of the normalization of the wave function forces ${\mathfrak C} = (2\pi)^{-1/2}$ (up to an overall and irrelevant phase). We have ignored the phase shift at turning points, since in our calculations below only a small portion of the orbit is considered.

In a multiple degree of freedom system that separates in the three coordinates $(x_1,x_2,x_3)$, Hamilton's characteristic function can be written for a given set of actions as a sum of functions in each separated variable. Then the total wave function is a product of the wave functions in each of the three coordinates, and Eq.~(\ref{eq:1dof}) generalizes to
\begin{equation}
\phi_{\bmath n}({\bmath x}) = (2\pi)^{-3/2} \sum_{\rm streams} \left| \frac{\partial x_1}{\partial\theta_1}
\frac{\partial x_2}{\partial\theta_2} \frac{\partial x_3}{\partial\theta_3} \right|^{-1/2}
\rme^{\rmi W({\bmath x},{\bmath n})/\hbar}.
\label{eq:WKB1}
\end{equation}
The summation is over the different streams, i.e. the different solutions for ${\bmath\theta}$ at fixed ${\bmath n}$ and ${\bmath x}$. In fact, Eq.~(\ref{eq:WKB1}) is even more general than that: if the wave equation separated in some other coordinate system $(X_1,X_2,X_3)$, then the transformation of a quantum mechanical wave function back to Cartesian coordinates $(x_1,x_2,x_3)$ introduces an additional factor of $|\det (\partial{\bmath X}/\partial{\bmath x})|^{1/2}$, leading to
\begin{equation}
\phi_{\bmath n}({\bmath x}) = (2\pi)^{-3/2} \sum_{\rm streams} \left| \det {\mathbfss T} \right|^{-1/2}
\rme^{\rmi W({\bmath x},{\bmath n})/\hbar},
\label{eq:WKB}
\end{equation}
where we have defined the $3\times 3$ matrix ${\mathbfss T} = (\partial x_a/\partial \theta_\mu)|_{\bmath n}$.
Therefore we may use Eq.~(\ref{eq:WKB}), even if the usual Cartesian coordinate system is not the system in which the motion separates.

For appropriate choice of photon modes, the basis functions can be taken to be local plane waves in the emitting region (which we place at the origin). An example of such a basis is the basis of spherical waves centered at a distant observer. Then
\begin{equation}
{\bmath Z}_\alpha({\bmath x};k) = Y_{\alpha}(k) \hat{\bmath\epsilon}_{\alpha}(k) \, \rme^{\rmi k\hat{\bmath s}_{\alpha}(k) \cdot{\bmath x}},
\end{equation}
where $Y_{\alpha,k}$ is a scalar amplitude, $\hat{\bmath\epsilon}_{\alpha}(k)$ is a unit vector polarization, and $\hat{\bmath s}_{\alpha}(k)$ is a unit vector in the direction of the local wave vector. The mode equation guarantees that this wave vector has norm $k$, and since the modes are transverse (divergenceless) we have $\hat{\bmath\epsilon}_{\alpha}(k)\perp\hat{\bmath s}_{\alpha}(k)$.

The interaction Hamiltonian, Eq.~(\ref{eq:HI}), is then
\begin{eqnarray}
\hat H^{\rm(I)}_{\rm int} 
\!\!\!\! &=& \!\!\!\! \frac{e}{4\pi^3} \sum_{{\bmath {nn}}'\alpha}\int \frac{\rmd k}{2\pi}\,
\sqrt{\frac{\pi c}{\hbar k\omega_{{\bmath n}}\omega_{{\bmath n}'} }} \hat a^\dagger_\alpha(k) \hat b_{\bmath n} \hat b^\dagger_{{\bmath n}'}
Y_{\alpha}^\ast(k) 
\int
 \rme^{-\rmi k\hat{\bmath s}_{\alpha}(k) \cdot{\bmath x}}
\nonumber \\ && \times
\frac{\hat{\bmath\epsilon}^\ast_{\alpha}(k)}{|\det{\mathbfss T}|}\cdot \Bigl[
-c\nabla_{\bmath x}W({\bmath x},{\bmath n}) + e{\bmath A}^{\rm(bg)}({\bmath x}) \Bigr]
\rme^{\rmi ({\bmath n}-{\bmath n}')\cdot{\bmath \theta}_{\rm cl}}
 \,\rmd^3{\bmath x},
\label{eq:HI-WKB}
\end{eqnarray}
where the difference in quantum numbers $\Delta{\bmath n} = {\bmath n}'-{\bmath n}$ is kept only in the relative phase of the different wave functions. Here ${\bmath\theta}_{\rm cl}({\bmath x},{\bmath n})$ is the classical angle at positon ${\bmath x}$ and action $\hbar{\bmath n}$ (i.e. obtained from the canonical transformation) and is equal to $\theta_\mu = \hbar^{-1} \partial W/\partial n_\mu$ by Eq.~(\ref{eq:direct1}). It appears in the complex exponential because the difference in $W$ between two different values of ${\bmath n}$ has been replaced by a partial derivative times $\Delta{\bmath n}$. We have suppressed here the sum over streams, as only one stream (the one beamed toward the observer) is significant at any one time.

The integrals in Eq.~(\ref{eq:HI-WKB}) can be simplified as follows. First, note that the classical velocity of the wave packet is
\begin{equation}
{\bmath u}_{\rm cl} = \nabla_{\bmath p} H_{\rm cl} = c\frac{c{\bmath p}_{\rm cl} - e{\bmath A}^{\rm(bg)}({\bmath x})}{H_{\rm cl}}
= c\frac{c\nabla_{\bmath x}W({\bmath x},{\bmath n}) - e{\bmath A}^{\rm(bg)}({\bmath x})}{H_{\rm cl}({\bmath n})}.
\end{equation}
We may thus write
\begin{equation}
\hat H^{\rm(I)}_{\rm int} 
= \hbar\sum_{\alpha,n,\Delta{\bmath n}}\int \frac{\rmd k}{2\pi}\,
 \hat a^\dagger_\alpha(k) 
\hat b^\dagger_{{\bmath n}+\Delta{\bmath n}} \hat b_{\bmath n}
{\mathfrak F}_{\alpha,{\bmath n},\Delta{\bmath n}}(k),
\label{eq:HG}
\end{equation}
where
\begin{equation}
{\mathfrak F}_{\alpha,{\bmath n},\Delta{\bmath n}}(k) = - \sqrt{\frac{4\pi c}{\hbar k}} Y_{\alpha}^\ast(k) F_{\alpha,{\bmath n},\Delta{\bmath n}}(k)
\label{eq:GF}
\end{equation}
and
\begin{equation}
F_{\alpha,{\bmath n},\Delta{\bmath n}}(k) = \frac ec\int \rme^{-\rmi k\hat{\bmath s}_{\alpha}(k) \cdot{\bmath x}}
\hat{\bmath\epsilon}^\ast_{\alpha}(k) \cdot {\bmath u}_{\rm cl}({\bmath x},{\bmath n})
\rme^{-\rmi\Delta{\bmath n}\cdot{\bmath \theta}_{\rm cl}}
 \,\frac{\rmd^3{\bmath\theta}}{(2\pi)^3}.
\label{eq:F}
\end{equation}
In this expression, we have used the Jacobian to convert the $\rmd^3{\bmath x}$ integral to a $\rmd^3{\bmath\theta}$ integral, and used that the classical Hamiltonian corresponding to a given orbit is equal to $\hbar\omega_{\bmath n}$.

The interpretation of $F_{\alpha,{\bmath n},\Delta{\bmath n}}(k)$ is straightforward. In {\em classical} electrodynamics, the current density ${\bmath J}_{\rm cl}({\bmath x},t)$ is multi-periodic with frequencies given by ${\bmath m}\cdot{\bmath\Omega}$, where ${\bmath m}\in{\mathbb Z}^3$ is a triplet of integers. The classical current density is given by $e{\bmath u}_{\rm cl}\delta^{(3)}({\bmath x}-{\bmath x}_{\rm cl})$, where ${\bmath x}_{\rm cl}$ is the particle's classical position.
Its 4-dimensional Fourier transform is
\begin{eqnarray}
\tilde{\bmath J}_{\rm cl}({\bmath k},\omega)
\!\!\!\! &=& \!\!\!\! \int {\bmath J}_{\rm cl}({\bmath x},t)\,\rme^{-\rmi({\bmath k}\cdot{\bmath x}-\omega t)}\, \rmd^3{\bmath x}\,\rmd t
\nonumber \\ &=& \!\!\!\! \int{\rmd^3\bmath\theta} \int \rmd t \, \delta^{(3)}[{\bmath\theta}-{\bmath\theta}(0) -{\bmath\Omega}t] \,e{\bmath u}_{\rm cl}\,\rme^{-\rmi{\bmath k}\cdot{\bmath x}_{\rm cl}} \,\rme^{\rmi\omega t}
\nonumber \\ &=& \!\!\!\! \sum_{\bmath m} \int \frac{\rmd^3\bmath\theta}{(2\pi)^3} \int \rmd t \, \rme^{\rmi{\bmath m}\cdot[{\bmath\theta}-{\bmath\theta}(0) -{\bmath\Omega}t]} \,e{\bmath u}_{\rm cl}\,\rme^{-\rmi{\bmath k}\cdot{\bmath x}_{\rm cl}} \,\rme^{\rmi\omega t}
\nonumber \\ &=& \!\!\!\!
\sum_{\bmath m} \int \frac{\rmd^3\bmath\theta}{(2\pi)^2}\, e{\bmath u}_{\rm cl}\, \rme^{-\rmi {\bmath k}\cdot{\bmath x}_{\rm cl}}
\rme^{\rmi{\bmath m}\cdot[{\bmath \theta}-{\bmath \theta}(0)]}
\delta(\omega-{\bmath m}\cdot{\bmath\Omega}),
\end{eqnarray}
where ${\bmath\theta}(0)$ is the angle at $t=0$.
The classical vector potential $\tilde{\bmath A}_{\rm cl,obs}(\omega)$ radiated by the particle, measured at a large distance $R$ from the system in direction $\hat{\bmath s}_\alpha$, is then given by
\begin{eqnarray}
&& \!\!\!\!\!\!\!\!\!\!\!\!\!\!\!\!
\hat{\bmath\epsilon}_\alpha^\ast\cdot\tilde{\bmath A}_{\rm cl,obs}(\omega)
\nonumber \\
&=& \!\!\!\!
\frac e{cR} \sum_{\bmath m} \int \frac{\rmd^3\bmath\theta}{(2\pi)^2}\, {\bmath\epsilon}_\alpha^\ast\cdot{\bmath u}_{\rm cl}\,
\rme^{\rmi k\omega/c}
\rme^{-\rmi k\hat{\bmath s}_\alpha\cdot{\bmath x}_{\rm cl}}
\rme^{\rmi{\bmath m}\cdot[{\bmath \theta}-{\bmath \theta}(0)]}
\delta(\omega-{\bmath m}\cdot{\bmath\Omega})
\nonumber \\
&=& \!\!\!\! \frac{\rme^{\rmi \omega R/c}}{R} \sum_{\bmath m} 2\pi F_{\alpha,{\bmath n},{\bmath m}}(\omega/c)
\rme^{-\rmi{\bmath m}\cdot{\bmath \theta}(0)}
\delta(\omega-{\bmath m}\cdot{\bmath\Omega}),
\label{eq:correspondence-o}
\end{eqnarray}
where in the intermediate steps we have taken $k=\omega/c$ and used the Green's function expansion for electromagnetic fields \citep[Eq. 6.48]{1998clel.book.....J} to equate the received frequency-domain magnetic vector potential to the 4-dimensional Fourier transform of the source. The conclusion is that the coefficient in the interaction Hamiltonian that contains the detailed dependence on the wave function properties is in fact equal to the classical radiated electromagnetic field. This is a manifestation of the correspondence principle.

The time-domain version of Eq.~(\ref{eq:correspondence-o}) is
\begin{equation}
\hat{\bmath\epsilon}_\alpha^\ast\cdot{\bmath A}_{\rm cl,obs}(t)
=\frac{\rme^{\rmi \omega R/c}}{R} \sum_{\bmath m} F_{\alpha,{\bmath n},{\bmath m}}(\omega/c)
\rme^{-\rmi{\bmath m}\cdot{\bmath \theta}(0)}
\rme^{-\rmi{\bmath m}\cdot{\bmath\Omega} t}.
\label{eq:correspondence-t}
\end{equation}
Note that $F_{\alpha,{\bmath n},{\bmath m}}(k)$ has units of magnetic flux.

\subsection{Quantum emission -- formalism}

Now let us follow the quantum behavior of the photon state through the emission of a single pulse, to second order in the interaction strength. In propagating from a starting time $t_{\rm s}$ to an ending time $t_{\rm e}$, the state varies as
\begin{eqnarray}
|\Psi(t_{\rm e})\rangle
\!\!\!\! &=& \!\!\!\!
\sum_{j=0}^\infty \frac{(-\rmi)^j}{\hbar^j}
\int_{t_{\rm e}>t_1>...t_j>t_{\rm s}}\rmd t_1...\rmd t_j\,
\hat U(t_{\rm e,1}) \hat H_{\rm int} \hat U(t_{1,2}) \hat H_{\rm int} ...
\nonumber \\ && \times
\hat U(t_{j-1,j}) \hat H_{\rm int} \hat U(t_{j,\rm s}) 
 |\Psi(t_{\rm s})\rangle
,
\end{eqnarray}
where $\hat U(\delta t) = \rme^{-\rmi (\hat H_{\rm rad}+\hat H_{\rm el})\delta t/\hbar}$. The shorthand $t_{\rm a,b} \equiv t_{\rm a}-t_{\rm b}$ has been introduced. One may define a ``rotated'' Hamiltonian\footnote{In the commonly used interaction picture formulation of quantum field theory, the interaction Hamiltonian operator is $\hat H_{\rm rot}(t)$ and the field operators evolve from one time to another according to $\hat U(\delta t)$.} by
\begin{equation}
\hat H_{\rm rot}(t) = \hat U(t_{\rm e}-t) \hat H_{\rm int} \hat U^\dagger(t_{\rm e}-t),
\end{equation}
with which
\begin{eqnarray}
|\Psi(t_{\rm e})\rangle
\!\!\!\! &=& \!\!\!\!
\sum_{j=0}^\infty \frac{(-\rmi)^j}{\hbar^j}
\int_{t_{\rm e}>t_1>...t_j>t_{\rm s}} \!\!\!\!\rmd t_1...\rmd t_j\,
\hat H_{\rm rot}(t_1) \hat H_{\rm rot}(t_2) ...  \hat H_{\rm rot}(t_j)
\nonumber \\ && \times
\hat U(t_{\rm es})|\Psi(t_{\rm s})\rangle.
\label{eq:temp-A1}
\end{eqnarray}
The ordering of time here is key, because while the interaction Hamiltonian $H_1$ always commutes with itself, it does not commute with the unperturbed Hamiltonian or hence with $\hat U(\delta t)$. It follows that the unequal-time rotated Hamiltonians do not necessarily commute with each other. However, by taking the logarithm of the operator on the first line of Eq.~(\ref{eq:temp-A1}), we find
\begin{equation}
|\Psi(t_{\rm e})\rangle
= \rme^{-\rmi \hat{\cal O}} \hat U(t_{\rm es})|\Psi(t_{\rm s})\rangle,
\label{eq:temp-A2}
\end{equation}
where
\begin{equation}
\hat{\cal O} = \frac1\hbar \int_{t_{\rm s}}^{t_{\rm e}} \rmd t_1\,\hat H_{\rm rot}(t_1)
- \frac{\rmi}{\hbar^2} \int_{t_{\rm s}}^{t_{\rm e}} \rmd t_1 \int_{t_{\rm s}}^{t_1} \rmd t_2\,
[\hat H_{\rm rot}(t_1),\hat H_{\rm rot}(t_2)]
 + ...
\end{equation}
is a Hermitian operator.

The terms in $\hat{\cal O}$ can be understood most easily if $H_{\rm rot}$ is broken down into individual terms (each with some number of annihilation and creation operators of particular modes) such that $H_{{\rm rot},a} \propto \rme^{-i\omega_{\rm a}t}$. Then -- taking the limit of large $t_{\rm es}$ -- the time integrals may be performed to give\footnote{This operator is easily seen to be Hermitian since each term $H_{{\rm rot},a}$ will have a conjugate term $H^\dagger_{{\rm rot},a}$ with the opposite frequency. In the second-order term in ${\cal O}$, the Hermitian conjugate term appears with a $-$ sign in addition to the usual complex conjugates since $1/(\omega_a-\omega_b)$ flips sign. This is why there is no factor of $\rmi$ in this term, even though for Hermitian operators $\hat A$ and $\hat B$ it is $\rmi[\hat A,\hat B]$ rather than $[\hat A,\hat B]$ is Hermitian.}
\begin{eqnarray}
\hat{\cal O} \!\!\!\! &=& \!\!\!\! \frac{t_{\rm es}}\hbar \sum_a {\cal W}(\omega_a) \hat H_{{\rm int},a}
+ \frac{t_{\rm es}}{\hbar^2} \sum_{a,b} {\cal W}(\omega_a+\omega_b)
{\mathbb P}\frac1{\omega_a-\omega_b}
\nonumber \\ && \times
[\hat H_{{\rm int},a},\hat H_{{\rm int},b}] + ...,
\label{eq:O-exp}
\end{eqnarray}
where ${\mathbb P}$ denotes a principal part\footnote{This is a principal part in the sense that one averages over the two possible pole displacements, ${\mathbb P}(z^{-1}) = \frac12[(z+\rmi\epsilon)^{-1} + (z-\rmi\epsilon)^{-1}]$. This way for an analytic function $f$, the conventional principal part of the integral $\int z^{-1} f(z)\,\rmd z$ is equal to $\int {\mathbb P}(z^{-1})\, f(z)\,\rmd z$.}, and ${\cal W}$ is a window function:
\begin{equation}
{\cal W}(s) = \frac1{t_{\rm es}}\int_{t_{\rm s}}^{t_{\rm e}} \rme^{\rmi s t_{\rm ea}}\,\rmd t_{\rm a} = \rme^{\rmi st_{\rm es}/2} \frac{\sin (st_{\rm es}/2)}{st_{\rm es}/2}
\end{equation}
with ${\cal W}(0)=1$, ${\cal W}(-s)={\cal W}^\ast(s)$, and $\int_{-\infty}^\infty {\cal W}(s)\,\rmd s=\pi/t_{\rm es}$.
To simplify the second-order term we used the identity that for long times\footnote{This can be proven by splitting the integral into terms symmetric under $\zeta\leftrightarrow\eta$ and antisymmetric. The symmetric term becomes the product of two ${\cal W}$-functions, while the antisymmetric term can be split into a double integral over $(t_{\rm a}+t_{\rm b})/2$ and $t_{\rm ab}$. Approximating the range of integration over $t_{\rm ab}$ as $0<t_{\rm ab}<\infty$ gives the result. The exact antisymmetry of this term under $\zeta\leftrightarrow\eta$ implies that the inverse $1/(\zeta-\eta)$ should be taken to be the principal part.},
\begin{equation}
\int_{t_{\rm s}}^{t_{\rm e}}\!\rmd t_{\rm a}\int_{t_{\rm s}}^{t_{\rm a}}\!\rmd t_{\rm b}\, \rme^{\rmi \zeta t_{\rm ea}}
\rme^{\rmi \eta t_{\rm eb}}
\approx \frac12t_{\rm es}^2 {\cal W}(\zeta) {\cal W}(\eta)
+2\rmi t_{\rm es}{\cal W}(\zeta+\eta) {\mathbb P}\frac1{\zeta-\eta}.
\end{equation}
The interpretation of Eq.~(\ref{eq:O-exp}) is straightforward: the long-time evolution is dominated by a series of interactions with 1 vertex, with 2 vertices, and higher-order terms (not shown here). Interactions with multiple vertices contain a propagator (inverse frequency denominator). This is the familiar expansion of particle scattering in quantum field theory.\footnote{The treatment of the propagator poles is different because here we do not have the boundary conditions of a scattering problem.} The exponential in Eq.~(\ref{eq:temp-A2}) allows multiple interactions to take place; it is of minor importance for single-particle scattering but is critical for understanding the coherence properties of light.

\subsection{Quantum emission -- application}
\label{as:de}

It is now time to consider the density matrix evolution of the radiation field during the above process. Suppose that we start in a coherent photon state and a definite action for the electron, i.e.
\begin{equation}
|\Psi(t_{\rm s})\rangle = |v_\alpha(k)\rangle \otimes |{\bmath n}\rangle.
\end{equation}
Here $|{\bmath n}\rangle$ indicates an electron in the state with quantum numbers ${\bmath n}$, i.e. $|{\bmath n}\rangle \equiv \hat b^\dagger_{\bmath n}|{\rm vac}\rangle$.
The unperturbed unitary evolution takes this to
\begin{equation}
|\Psi^{(0)}_{\rm e}\rangle = \rme^{-\rmi\omega_{\bmath n}t_{\rm es}}|\rme^{-\rmi ckt_{\rm es}} v_\alpha(k)\rangle \otimes |{\bmath n}\rangle.
\end{equation}
The final photon density matrix is
\begin{equation}
\rho_{\rm rad}(t_{\rm e}) = {\rm Tr}_{\rm el} \Bigl[
\rme^{-\rmi\hat {\cal O}} 
|\rme^{-\rmi ckt_{\rm es}} v_\alpha(k)\rangle \otimes |{\bmath n}\rangle
\langle\rme^{-\rmi ckt_{\rm es}} v_\alpha(k)| \otimes \langle{\bmath n}| \rme^{\rmi\hat{\cal O}}
\Bigr],
\end{equation}
where the trace is over the electron state. The trace may be simplified with a resolution of the identity operator into angle states as
\begin{equation}
{\mathbb I}_{\rm el} = \int \frac{\rmd^3\bmath\vartheta}{(2\pi)^3}\,|{\bmath\vartheta}\rangle\langle{\bmath\vartheta}|,
\end{equation}
where $|{\bmath\vartheta} \rangle \equiv \sum_{\bmath m} \rme^{-\rmi{\bmath\vartheta}\cdot{\bmath m}} |{\bmath m}\rangle$.
Then
\begin{equation}
\rho_{\rm rad}(t_{\rm e}) = 
\int \frac{\rmd^3\bmath\vartheta}{(2\pi)^3}\,
\langle {\bmath\vartheta}|\rme^{-\rmi\hat {\cal O}} 
|\rme^{-\rmi ckt_{\rm es}} v_\alpha(k)\rangle \otimes
 |{\bmath n}\rangle
\langle\rme^{-\rmi ckt_{\rm es}} v_\alpha(k)| \otimes \langle{\bmath n}| \rme^{\rmi\hat{\cal O}}
|{\bmath\vartheta}\rangle.
\label{eq:rad1}
\end{equation}

Now consider the effect of the terms in $\hat{\cal O}$ that are first-order in $e$, which correspond to the elementary emission and absorption processes. We now make the approximation that, over a small range in electron quantum numbers near ${\bmath n}$, the amplitude $F_{\beta,{\bmath n}',\Delta{\bmath n}}(k')$ is roughly constant. This assumption eliminates self-absorption, since self-absorption with the absorbing electron in a given level $n$ is related to the fact that  transitions from $n\leftrightarrow n+1$ have a stronger oscillator strength than transitions from $n\leftrightarrow n-1$.\footnote{It can be seen that this situation will occur semi-classically from Eq.~(\ref{eq:correspondence-t}). Considering only one of the electron degrees of freedom  and assuming a harmonic oscillator, the squared amplitude of emitted radiation is proportional to the action, $|F_{\beta,n',1}(k')|^2\propto n'$.} Then $\hat{\cal O}$ reduces to
\begin{equation}
\hat{\cal O} \approx t_{\rm es} \sum_{\beta,\Delta{\bmath n}} \int \frac{\rmd k'}{2\pi}
\,{\cal W}(ck'-{\bmath\Omega}\cdot\Delta{\bmath n}) {\cal F}_{\beta,{\bmath n},\Delta{\bmath n}}\hat\Sigma^\dagger_{\Delta{\bmath n}}
\hat a_\beta^\dagger(k')
+ {\rm h.c.},
\end{equation}
where $\hat\Sigma_{\Delta{\bmath n}} = \sum_{{\bmath m}} \hat b^\dagger_{\bmath m} \hat b_{{\bmath m}+\Delta{\bmath n}}$ is the state shift operator. Given our previous approximations, this is now the only operator in $\hat{\cal O}$ that acts on the electron Hilbert space. But in Eq.~(\ref{eq:rad1}), it acts on an angle state, which is an eigenstate\footnote{Technically this is only true if the range of quantum numbers is over all integers, since otherwise the eigenstate formula fails for e.g. $n_1<\Delta n_1$. Since our analysis does not involve states with small quantum numbers, this is not a problem.}:
\begin{equation}
\hat\Sigma_{\Delta{\bmath n}} |{\bmath\vartheta}\rangle = \rme^{-\rmi{\bmath\vartheta}\cdot\Delta{\bmath n}}|{\bmath\vartheta}\rangle
~~~{\rm and}~~~
\hat\Sigma^\dagger_{\Delta{\bmath n}} |{\bmath\vartheta}\rangle = \rme^{\rmi{\bmath\vartheta}\cdot\Delta{\bmath n}}|{\bmath\vartheta}\rangle.
\end{equation}
We may thus make the replacement in Eq.~(\ref{eq:rad1}):
\begin{equation}
\hat{\cal O} \approx t_{\rm es} \sum_{\beta,\Delta{\bmath n}} \int \frac{\rmd k'}{2\pi}
\,{\cal W}(ck'-{\bmath\Omega}\cdot\Delta{\bmath n}) {\cal F}_{\beta,{\bmath n},\Delta{\bmath n}}
\rme^{\rmi{\bmath\vartheta}\cdot\Delta{\bmath n}} \hat a_\beta^\dagger(k')
+ {\rm h.c.},
\end{equation}
and use $\langle{\bmath n}|{\bmath\vartheta}\rangle = \rme^{-\rmi{\bmath\vartheta}\cdot{\bmath n}}$. This leaves Eq.~(\ref{eq:rad1}) in the form
\begin{equation}
\rho_{\rm rad}(t_{\rm e}) = 
\int \frac{\rmd^3\bmath\vartheta}{(2\pi)^3}\,
\hat{\cal D} |\rme^{-\rmi ckt_{\rm es}} v_\alpha(k)\rangle 
\langle\rme^{-\rmi ckt_{\rm es}} v_\alpha(k)| \hat{\cal D}^\dagger,
\label{eq:rad2}
\end{equation}
where
\begin{equation}
\hat{\cal D} = \exp \Bigl\{
\sum_\beta \int \frac{\rmd k'}{2\pi} [u_\beta(k') \hat a^\dagger_\beta(k') - u_\beta^\ast(k') \hat a_\beta(k') ]
\Bigr\}
\end{equation}
and
\begin{equation}
u_\beta(k') = -\rmi t_{\rm es} \sum_{\Delta{\bmath n}}
{\cal W}(ck'-{\bmath\Omega}\cdot\Delta{\bmath n}) {\cal F}_{\beta,{\bmath n},\Delta{\bmath n}}
\rme^{\rmi{\bmath\vartheta}\cdot\Delta{\bmath n}}.
\label{eq:ub}
\end{equation}
Note that $\hat{\cal D}$ is a displacement operator; acting on a coherent state $|v_\alpha(k)\rangle$, it gives another state $\rme^{\rmi\chi}|u_\alpha(k)+v_\alpha(k)\rangle$, whose amplitude is the input amplitude plus the displacement $u_\alpha(k)$. The phase $\chi$ is not needed here, since it cancels out in the density matrix Eq.~(\ref{eq:rad2}).

The result of Eq.~(\ref{eq:rad2}) is that the output state of the radiation, after interaction with a single electron in a quantum state $|{\bmath n}\rangle$, is a statistical superposition of coherent states, where the statistical average is taken over angles ${\bmath\vartheta}$. The coherent state is displaced by $u_\alpha(k)$, given by Eq.~(\ref{eq:ub}).

The above machinery is now well-suited to studying the quantum state of the radiation after interaction with {\em many} electrons starting from an initial vacuum state $|{\rm vac}\rangle$. The interaction with each electron adds another term to $u_\alpha(k)$, thus placing the photon ultimately in a statistical superposition of coherent states with amplitude $\sum_{i=1}^{N_e} u_\alpha(k)$, where the statistical superposition is taken over the $3N_e$ angles ${\bmath\vartheta}_1 ... {\bmath\vartheta}_{N_e}$:
\begin{equation}
\rho_{\rm rad}(t_{\rm e}) = \int \frac{\rmd^3{\bmath\vartheta}_1 ... \rmd^3{\bmath\vartheta}_{N_e}}{(2\pi)^{3N_e}}\,
\left| \sum_{i=1}^{N_e} u_\alpha(k) \right\rangle \left\langle \sum_{i=1}^{N_e} u_\alpha(k) \right|.
\label{eq:rad3}
\end{equation}

\subsection{Relation to classical waveform}

The interpretation of this result is easiest if we realize that to every coherent state $|u_\alpha(k)\rangle$ of a quantum field there corresponds a classical field configuration obtained via the substitution $\hat a_\alpha(k)\rightarrow u_\alpha(k)$ in Eq.~(\ref{eq:Arad}). The quantum state is the vacuum displaced by this classical solution, and as such a measurement on a quantum system in a nonnegative-weight statistical superposition of coherent states (e.g. Eq.~\ref{eq:rad3}) can contain no more information than the classical system in the corresponding statistical distribution. In particular, according to the optical equivalence theorem \citep{1963PhRvL..10..277S}, the normal-ordered quantum correlation functions of the field, such as the intensity fluctuations that would be measured by a photoelectric detector, are equal to the classically computed moments based on the probability distribution of $u_\alpha(k)$.

We are therefore motivated to learn about the functions $u_\alpha(k)$ produced in single-electron interactions and their statistical properties. We do this by finding the corresponding received field at the observer. Substituting ${\cal W}(s) \rightarrow (\pi/t_{\rm es})\delta(s)$ into Eq.~(\ref{eq:ub}), we find
\begin{equation}
u_\alpha(k) = \rmi\pi \sum_{\Delta\bmath n}^+ \delta(ck-{\bmath\Omega}\cdot\Delta{\bmath n})
\sqrt{\frac{4\pi c}{\hbar k}}\,Y^\ast_\alpha(k) F_{\alpha,{\bmath n},\Delta{\bmath n}}(k) 
\rme^{\rmi{\bmath\vartheta}\cdot\Delta{\bmath n}},
\end{equation}
where the $+$ sign on the summation indicates that the sum is taken over states with ${\bmath\Omega}\cdot\Delta{\bmath n}>0$.
Substitution into Eq.~(\ref{eq:Arad}) with then yields, with some simplification, the classical-equivalent field
\begin{equation}
{\bmath A}_{\rm coh}({\bmath x}) = 2\pi\rmi \sum_{\alpha,\Delta{\bmath n}}^+ Y^\ast_\alpha(k) {\bmath Z}_\alpha({\bmath x};k)
\frac{F_{\alpha,{\bmath n},\Delta{\bmath n}}(k)}{k} \rme^{\rmi {\bmath\vartheta}\cdot\Delta{\bmath n}}
+ {\rm c.c.},
\end{equation}
where here $k=\omega/c$ and $\omega = {\bmath\Omega}\cdot\Delta{\bmath n}$.
The final step is the evaluation of the mode functions. Let us take a set of modes propagating near the direction $\hat{\bmath s}$, which will be taken to be toward the observer, and take the polarization vectors to be either horizontal or vertical. Then the behaviour of the modes near the source (origin) will be that one mode is a top-hat with cross sectional area ${\mathfrak A}$ (taken to be large compared to the emitting region): then near the origin
\begin{equation}
{\bmath Z}_\alpha({\bmath x};k) = \frac1{\sqrt{\mathfrak A}} \hat{\bmath\epsilon}_\alpha \rme^{\rmi k \hat{\bmath s}\cdot{\bmath x}}
\end{equation}
within the area ${\mathfrak A}$ and 0 otherwise, so that $Y_\alpha(k) = {\mathfrak A}^{-1/2}$. Aside from these 2 modes (2 since there are both polarizations), the remaining photon modes do not contribute. The Kirchhoff diffraction formula \citep[Eq. 10.85]{1998clel.book.....J} then gives 
${\bmath Z}_\alpha({\bmath x};k)$ at the observer by integrating over the area ${\mathfrak A}$,
\begin{equation}
{\bmath Z}_\alpha({\bmath x}_{\rm obs};k) = {\mathfrak A}^{1/2} \hat{\bmath\epsilon}_\alpha \frac{k \rme^{\rmi kR}}{2\pi\rmi R}.
\end{equation}
Thus:
\begin{equation}
{\bmath A}_{\rm coh}({\bmath x}) = \sum_{\alpha,\Delta{\bmath n}}^+ \frac{\rme^{\rmi kR}}{R}
\hat{\bmath\epsilon}_\alpha
F_{\alpha,{\bmath n},\Delta{\bmath n}}(k) \rme^{\rmi {\bmath\vartheta}\cdot\Delta{\bmath n}}
+ {\rm c.c.}.
\end{equation}
This is equivalent to Eq.~(\ref{eq:correspondence-t}) with relabeled phase factors, showing that {\em the quantum emission process from an optically thin electron cloud leads to a statistical superposition of coherent radiation states, with amplitude given by the classical field configuration, and statistical weight uniformly distributed over the classical angles (phases) of the electron trajectories}.

In the case where dispersion is present, each $k$-oscillator remains a harmonic oscillator but the frequency changes adiabatically as the wave enters and exits an ionized cloud with continuous electron density. Since the time evolution of a coherent state in a harmonic oscillator (Hamiltonian proportional to $\hat a^\dagger\hat a$) is that the complex amplitude $v$ picks up a phase, the cloud changes a coherent state $|v_\alpha(k)\rangle$ to a re-phased coherent state $|\rme^{-\rmi\phi(k)}v_\alpha(k)\rangle$. Again the effect is exactly as in classical physics, except that it acts on a coherent state displacement rather than a classical complex number \citep{1966PhL....21..650G}.

\end{document}